\documentclass[manuscript]{aastex}

\slugcomment{}

\shorttitle{}
\shortauthors{}

\begin{document}

\title{Sample variance in N--body simulations and \\
impact on tomographic shear predictions}


\author{Luciano Casarini \altaffilmark{1,2}, Oliver~F. Piattella \altaffilmark{1,2}}
\affil{Federal University of Espirito Santo (UFES), Department of Physics, Vitoria ES, Brazil}
\email{casarini.astro@gmail.com, oliver.piattella@pq.cnpq.br}

\and

\author{Silvio~A. Bonometto \altaffilmark{3}, Marino Mezzetti \altaffilmark{3}} 
\affil{University of Trieste, Deparment of Physics, Astronomy Unit, Trieste (TS), Italy}
\email{bonometto@oats.inaf.it, mezzetti@oats.inaf.it}

\altaffiltext{1}{National Council for Scientific and Technological Development (CNPq), Brazil}
\altaffiltext{2}{Espirito Santo Research Foundation (FAPES), Brazil}
\altaffiltext{3}{National Institute for Astrophysics (INAF), Italy}

\begin{abstract}

We study the effects of sample variance in N--body simulations, as a 
function of the size of the simulation box, namely in connection 
with predictions on tomographic shear spectra. We make use of a set 
of 8 $\Lambda$CDM simulations in boxes of 128, 256, 512 $h^{-1}$Mpc 
aside, for a total of 24, differing just by the initial seeds. Among 
the simulations with 128 and 512 $h^{-1}$Mpc aside, we suitably 
select those closest and farthest from {\it average}. Numerical and 
linear spectra $P(k,z)$ are suitably connected at low $k$ so to 
evaluate the effects of sample variance on shear spectra 
$C_{ij}(\ell)$ for 5 or 10 tomographic bands. We find that shear 
spectra obtained by using 128 $h^{-1}$Mpc simulations can vary up to 
$\sim 25\, \%$, just because of the seed.  Sample variance lowers to 
$\sim 3.3\, \%$, when using 512 $h^{-1}$Mpc. These very 
percentages could however slightly vary,  if other sets of the same 
number of realizations were considered. Accordingly, in order to 
match the $\sim 1\, \%$ precision expected for data, if still using 
8 boxes, we require a size $\sim  1300$ --$ 1700 \, h^{-1}$ 
Mpc for them.

\end{abstract}


\section{Introduction}
The tidal gravitational field of density inhomogeneities distorts the
images of distant galaxies in the Universe. This effect, dubbed {\it
  cosmic shear}, was first observed in 2000, by correlating distant
galaxy ellipticities \citep{Bacon et al. 2000, Kaiser et al. 2000, van
  Waerbeke et al. 2000, Wittman et al. 2000}. In turn, comparing
cosmic shear with model prediction is expected to become a critical
pattern for model selection, namely if data are suitably shared in
redshift bands, so to create a sort of cosmic {\it tomography}. The
key point being that tidal fields allow us a more direct insight into
the distribution of masses, independently of light emission mechanisms.

The significance of cosmic shear data however goes even beyond that. 
Being obtained from low--$z$ systems, they are indeed complementary 
to high--$z$ CMB anisotropy measurements. Furthermore, in respect to 
other low--$z$ observables, as SNIa redshift distributions, cosmic 
shear exhibits a specific dependence on the dynamics of structure 
growth (e.g. \cite{Hu 2002, Albrecht et al. 2006, Peacock et al. 
2006, La Vacca and Colombo 2008}), so enabling us to test the 
consistency between background and inhomogeneity evolutions.

It is therefore hardly surprising that a number of experiments have
been planned as, e.g.,
BOSS\footnote{http://www.sdss3.org/surveys/boss.php},
PanStarrs\footnote{http://pan-starrs.ifa.hawai.edu},
HETDEX\footnote{http://hetdex.org/hetdex},
DES\footnote{http://www.darkenergysurvey.org},
LSST\footnote{http://www.lsst.org}, 
KIDS\footnote{http://kids.strw.leidenuniv.nl}
WFIRST\footnote{http://wfirst.gsfc.nasa.gov},
and
Euclid\footnote{http://www.euclid-ec.org/‎} \citep{Amendola et al. 2013},
aiming to scan a large area of the sky, seeking fresh information on
weak lensing.

The necessary tool to exploit this information are predictions on the
distribution of density inhomogeneities and their evolution, both on
linear and non–linear scales. The former predictions can be obtained
through library algorithms, like CAMB \citep{camb}. The latter ones,
on the contrary, can only be based on simulations and therefore depend
on the initial realization of matter distribution through point
particles.

Such dependence gradually attenuates when greater cosmic volumes are
simulated. Accordingly, the Millennium simulations of a $\Lambda$CDM
cosmology \citep{millenium} were performed in a box of 
500$\, h^{-1}$Mpc aside. Another, more recent, large (in russian
``Bolshoi'') simulation of $\Lambda$CDM \citep{bolshoi} was run by
using 2048$^3$ particles, although in a 250$\, h^{-1}$Mpc box. Then, the Deus
simulation series \citep{deus} of RP \citep{rp} and SUGRA \citep{sugra} 
cosmologies (besides of $\Lambda$CDM) were run in boxes of
quite large sizes, up to 1296$\, h^{-1}$Mpc. Let us finally mention
the significant systematic effort deployed to create the Coyote
Universe simulation suite \citep{Heitmann et al. 2010, Heitmann et al. 2013}.  It
yields a prediction scheme for the matter power spectrum (the
so-called {\it emulator}), accurate at the 1$\, \%$ level, out to
$k\sim 1\, $Mpc$^{-1}$ and redshift z=1. Their simulations were run in
boxes with side length up to 1300 Mpc and tested a wide set of wCDM
cosmologies with constant $w$ comprised between -0.7 and -1.3~.

Let us outline that, by using the technique introduced by
\cite{Casarini et al. 2009}, the Coyote Universe emulator can be also
exploited to find spectra of variable--$w$ cosmologies. Limitations on
$z$ are then more severe as, at any $z$, the technique requires
information on a range of constant--$w$ spectra wider than those
considered at $z=0$ and selected in a non--trivial way (clearly, if a
model is characterized by a $z$--dependent state equation yielding
$w(z) = \bar w$, the simulation better approaching the model at that
$z$ is not the one with $w = \bar w~).$

Among large simulations, it is also worth mentioning the recent
Millennium--XXL simulation, dealing with a $\Lambda$CDM model only,
but using a box of 3$\, h^{-1}$Gpc aside (Angulo \& White 2011).

{\rm A basic question that shear experiment will enable us to test,
  however, is whether one of these {\it classical} cosmologies is
  sufficient to approach data. In this case we expect full consistency
  between background and inhomogeneity evolutions. Any doubt on that
  would be an evidence of GR (General Relativity) violations
  \citep{Capozziello et al. 2006, Amendola et al. 2007} or energy
  exchanges between dark cosmic components \citep{Ellis et al. 1989,
    Wetterich 1995, Amendola2000, Amendola2002, Amendola2003, Klypin et al. 2003,
    Amendola2004, Das et al. 2005}, not to tell about even more
  complex options. Although N--body programs were built to tackle a 
  number of these options \citep{Maccio' et al. 2004, Baldi et al. 
  2010, Puchwein et al. 2013}, no available simulation set enables 
  us to afford immediate tests, while it would also be hard to 
  provide, {\it a priori}, a sufficiently wide range of simulations, 
  to cover a significant set of alternatives.}

Accordingly, it is quite relevant to test how far {\it sample
  variance} may affect model predictions trying to meet future data.
Of course, sample variance is no obstacle to comparing different
cosmologies on a theoretical basis: one just needs to start from the
same seed for all models. But, when trying to discriminate between
models through shear observations, one must make sure that sample
variance, 
within the simulation sample prepared to 
build angular spectra, stands well below model discrepancies.

This paper is therefore dedicated to test how sample variance depends
on the box size, as well as how much it can be set under control by
using a set of simulations in equal boxes, starting from different
seeds. 
More precisely, we aim to test how sample variance between
  simulation boxes is transfered into tomographic shear spectra.
Being obtainable by integrating through the former ones, their
variance can be expected to decrease; how strong this variance damping
can be can only be inspected through direct tests.  To do all that it
is however adequate to work within the context of $\Lambda$CDM models.
At the basis of this work there are therefore sets of 8 $\Lambda$CDM
simulations, in boxes with $L = 128$, 256, 512 $h^{-1}$Mpc aside, for
a total of 24.

 Before these predictions are directly applied to a specific
  experiment, however, one should apply the related survey window.
  For instance, non--linear mode coupling effects could depend on
  detailed observational features (see, e.g., \cite{Hamilton}). Such a
  detailed study, however, goes beyond the scope of the present
  analysis.

The observable we shall deduce from our simulations, first of all, are
the fluctuation spectra $P(k,z)$. We shall then use them to predict
shear spectra $C_{ij}(\ell)$, with 5 or 10 tomographic bands, labelled
by $i,\, j$.  This will then allow us to test if (and how) the sample
variance in shear spectra is related to the number of tomographic
bands.

To perform our 24 N--body simulations, we choose model parameters
consistent with recent Planck outputs \citep{Planck collaboration 2013},
shown in Table I, where symbols bear their usual meaning:
\vglue .2truecm
\centerline{Table I}
\vglue .01truecm
\centerline{ -------------------------------------------- }
\vglue -1.truecm
$$
\matrix{ 
h    & \Omega_b & \Omega_c &  n_s & \sigma_8 \cr
0.69 & 0.048    &  0.249   & 0.966 & 0.82 
}      
$$
\vglue -.1truecm
\centerline{ -------------------------------------------- } 

The N--body program used is {\sc PDKGRAV} \citep{pkdgrav}. Initial
conditions were produced with {\sc graphic–2} \citep{grafic2} and,
therefore, did not try to account for the impact of wavelength above
the box side (the so--called DC modes; see, e.g., \cite{Sirko 2005, Li
  et al 2014}) which, in our case and as will be verified, would yield
substantially unappreciable corrections. The particle numbers,
proportional to the box volume, are 128$^3$, 256$^3$, 512$^3$,
respectively. In all simulations $k_{Nyq} = \pi \, h\, {\rm Mpc}^{-1}$.

Simulation outputs are then provided for a large number of redshifts
$z_i$. Between z = 0 and z = 0.1 outputs stand at a redshift distance
$\Delta z = 0.01\, $. Then: between $z = 0.1$ and $z = 1$, $\Delta z =
0.1\, ;$ between z = 1 and z = 3, $\Delta z = 0.2\, ;$ finally,
outputs were obtained at a distance $\Delta z = 1$ up to z = 10$\, .$

The plan of the paper is therefore as follows. In Section 2 we shall
discuss the relation between number of realizations and sample
variance spanned. In Section 3 we shall then deepen another essential
question: how spectral points deduced from simulations can be
interpolated with linear spectra, at low $k$; and, which pattern shall
be followed to use non--linear results at large $k$'s, when numerical
noise and/or lack of resolution hide the numerical signal. These
problems were often overlooked in the literature; their technical
solutions are an original aspect of this analysis. Let us soon
outline, however, that our choices were also aimed to avoid artificial
differences between realizations; e.g., at large $k$'s one could
surely achieve better results, if this aim is disregarded. In Section
4 we shall debate the formation of tomographic filters, both for shear
and intrinsic spectra. In Section 5 and 6, we shall give the
expressions for shear and intrinsic spectra, and exhibit them for one
of the relevant cases, both putting in evidence the impact of
intrinsic deformations and determining which tomographic spectra are
essentially clean from this kind of contamination. The residual contamination,
will be taken as a meter for the residual sample variance we may allow
for. The final Section is devoted to a general discussion and to
drawing our conclusions.

\section{Number of realizations \& sample variance}
In this work we run $N_R=8$ simulations of a fixed $\Lambda$CDM model
for each box size considered. Such $N_R$ simulations differ just by
the pseudo-random number seed (just ``seed'', in the sequel) used by
{\sc graphic--2} to create initial conditions at $z_i=50$. These
differences, magnified at lower redshift, mimic the observational
discrepancies between real cosmic volumes of the same size, yielding
the so--called {\it sample variance}. This Section tries to predict
how much sample variance is spanned by $N_R$ realization.

The estimator used in this Paper to derive power spectra from the
simulations is described at the beginning of next Section. The
averaging procedure described there allows us to assume that, thanks
to the Central Limit theorem, discrepancies, at any given $k$ value,
are normally distributed.

So, let us suppose to use the $N_R$ particle
distributions to determine the fluctuation spectra $P_n(k,z)$
($n=1,2,...,N_R$), at each $z$. A further $(N_R+1)$--th realization,
in general, might yield spectra $P_{N_R+1}(k,z)$ laying among the
previous $N_R$ realizations, or widening their functional space. For
greater $N_R$, of course, the probability of keeping within the space
spanned by the $N_R$ initial spectra increases. Once $N_R$ is fixed,
however, what is the probability that $P_{N_R+1}(k,z)$ lays among the
former $P_n(k,z)$ spectra$\, $?

This is a hard and somehow ambiguous question; e.g., we should detail
when $P_{n+1}(k,z)$ is considered to lay among the previous $P_n(k,z)$
($n=1,2,...,N_R$) spectra and how the $P(k,z)$ distribution is far
from normal.

Here we shall therefore regress to a simpler question. At a given $z$,
let us then take a generic wave number $\bar k$ and assume that the
$N_{R}$ values of the power spectrum $P_n(\bar k) \equiv \phi_n$ are
randomly drawn from a normal distribution (see above). Let then $\mu$
be the mean of such $N_R$ values $\phi_n$ and let $\varepsilon$ be the
modulus of the maximum deviation --positive or negative-- of $\phi_n$
from $\mu$.  In this context, we try to evaluate the probability
$p_{N_{R}} (>\varepsilon)$ that a further value
$\phi_{N_R+1}=P_{N_R+1}(\bar k)$ differs from $\mu$ more than
$\varepsilon$ (let us draw the reader's attention on the difference
between probabilities $p$ and spectra $P$, indicated with small and
capital letters, respectively). This will be assumed to approach the
probability that a $N_R+1$--th spectrum has to lay among the previous
$N_R$.
\begin{figure}
\vskip -2.truecm
\epsscale{.5}
\plotone{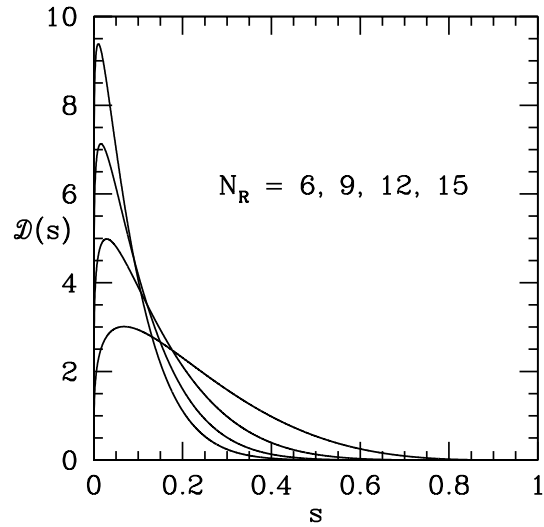}
\vskip -2.5truecm
\caption{Numerically observed distributions, normalized to unity, of
  $s$ values.  Greater $N_R$'s yield more peaked $\cal D$. \label{G6} }
\end{figure} 
\begin{figure}
\epsscale{.5}
\vskip -3.truecm
\plotone{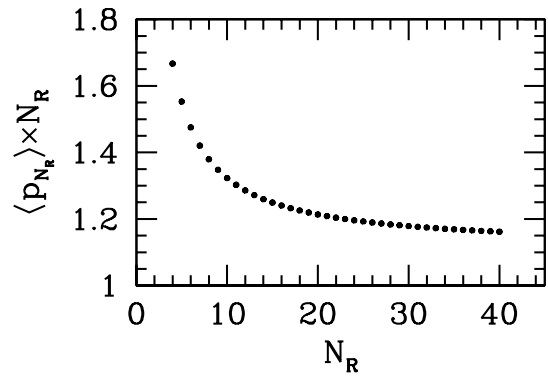}
\vskip -2.truecm
\caption{Expected probability that the fluctuation spectrum for an
  $(N_R+1)$--th realization lays significantly outside the range of
  spectra obtained from $N_R$ previous realization. Values of power
  spectrum, at any given $k$ value, are assumed to be normally
  distributed about their average.
\label{nr} }
\end{figure}

We can derive a first estimate of the probability of a new $P(k)$
having a value outside the range of current samples by means of the
quantiles of the Normal distribution.
More in detail, we assume that $\mu$ coincides with the peak of a
suitable normal distribution $G(s)$, and that the unit area below the
Gaussian curve is shared in $N_R=8$ equal parts, so that $N_R/2=4$
values $\phi_n$ lay at each side of $\mu$. In the ideal case, the most
distant $\phi_n$ shares then in two equal parts the 4--th (most
distant) area, so that the part of area beyond the most distant
$\phi_n$ holds $1/(2N_R) = 1/16 \, $, at each side of the
distribution. Summing up both sides, the area is 12.5$\, \%$ of the
total normalized area.

Let us soon outline that, if $N_R$ is small, $\mu$ hardly
coincides with the peak of the distribution, while each $\phi_n$ value
hardly shares in two equal parts its expected interval. This
admittedly rough argument however allows us an estimate.

In order to obtain a more reliable estimate of the expected
probability
\begin{equation}
p_{N_{R}}(>\varepsilon) = 1-{\cal G}(\mu+\varepsilon)+{\cal
  G}(\mu-\varepsilon)
\end{equation}
we make a large number ($10^{6}$) of random replicas of $N_R$ values.
In the ``ideal case'' described above, $p_{N_{R}}$ is 0.125~. After
each random replica, however, we can directly measure the value to be
taken, instead of 0.125, and our large number of replicas allow us to
determine the frequency distribution of each $s=p_{N_R}(>\varepsilon)$
value. In this way we find that such distribution holds the shape
\begin{equation}
f(s) \propto (1-s)^{1/n}\cdot s^{m}\, .
\end{equation} 
The values
\begin{equation}
\label {nm}
n = 0.5 \times (N_R-2) + 1.8~,~~
m = 0.9 \times (N_R-2)
\end{equation}
yield and excellent approximation to the observed distributions, for
$N_R > 4$. They are also shown in Figure \ref{G6}, normalized to unit
area, for $N_{R}=\:6,\:9,\:12,\:15$.

Accordingly, the expected probability
\begin{equation}
\langle p_{N_{R}}(>\varepsilon) \rangle = {\int_0^1ds ~s\, (1-s)^{1/n}s^m
\over \int_0^1 ds ~(1-s)^{1/n}s^m}~,
\end{equation}
while, quite in general, it is
$$
\int_0^1 ds~s^{\nu-1}(1-s)^{\mu-1} = B(\mu,\nu) = {\Gamma(\mu)\Gamma(\nu)
\over \Gamma(\mu+\nu)}~;
$$
here $B$ is the $\beta$--function, while the $\Gamma$--functions are
the analytic extensions of factorials\footnote{I.S.~Gradshteyn \& 
I.M.~Ryzhik, Academic Press, 1980 edition: 3.191.3, 8.384.1~.}.
Owing to the relation $\Gamma(s+1) = (s+1) \Gamma(s)$ it is then
immediate to obtain that
\begin{equation}
\langle p_{N_{R}}(>\varepsilon) \rangle = {1/n+1 \over 1/n+m+2}
\end{equation}
For $N_R=8$, we then obtain an expected probability of $15.88\, \%$, 
The complementary probability that $\phi_{N_R+1}$ falls inside the
interval $\mu \pm \epsilon$, approximately spanned by the first $N_R$
values, for $N_R=8$, is then $\sim 84\, \%$, approximately
corresponding to 1.5~standard deviations.

The conclusion that the spectra obtained from a set of 8 equal
simulations cover, approximately, $1.5\, \sigma$'s in the space of
possible spectra, seems therefore a reasonable estimate. For a generic
value of $N_R$, the ratio $\langle p_{N_R} \rangle/N_R^{-1}$ is shown
in Figure \ref{nr}. Even for the largest $N_R$ values considered here,
such ratio still keeps significantly above unity.

\section{Fluctuation spectra}
Fluctuation spectra as those considered in the previous Section were
obtained from simulations by using the algorithm PMpowerM included in
the PM package \citep{Klypin and Holtzman 1997}. Through a CiC
procedure, the algorithm assigns the density field on a uniform
cartesian grid starting from the particle distribution.

Here we consider effective values $n =2^f N$ ($N=128$,~256,~512), with
$f$ from 0 to 3; e.g., for $N=128$ (512) simulations we arrive to
$n=1024$ (4096). As is known, such large $n$ are obtainable by
considering a $N^3$ grid in a box of side $L / 2^f$, where all
simulation particles are inset, in points of coordinate $x_{i,f} = x_i
- \nu L/2^f$, $\nu$ being the smallest integer number allowing $0 <
x_{i,f} < L/2^f$ ($i=1,2,3$ labels spatial coordinates).

\begin{figure}
\epsscale{.6}
\vskip -.9truecm
\plotone{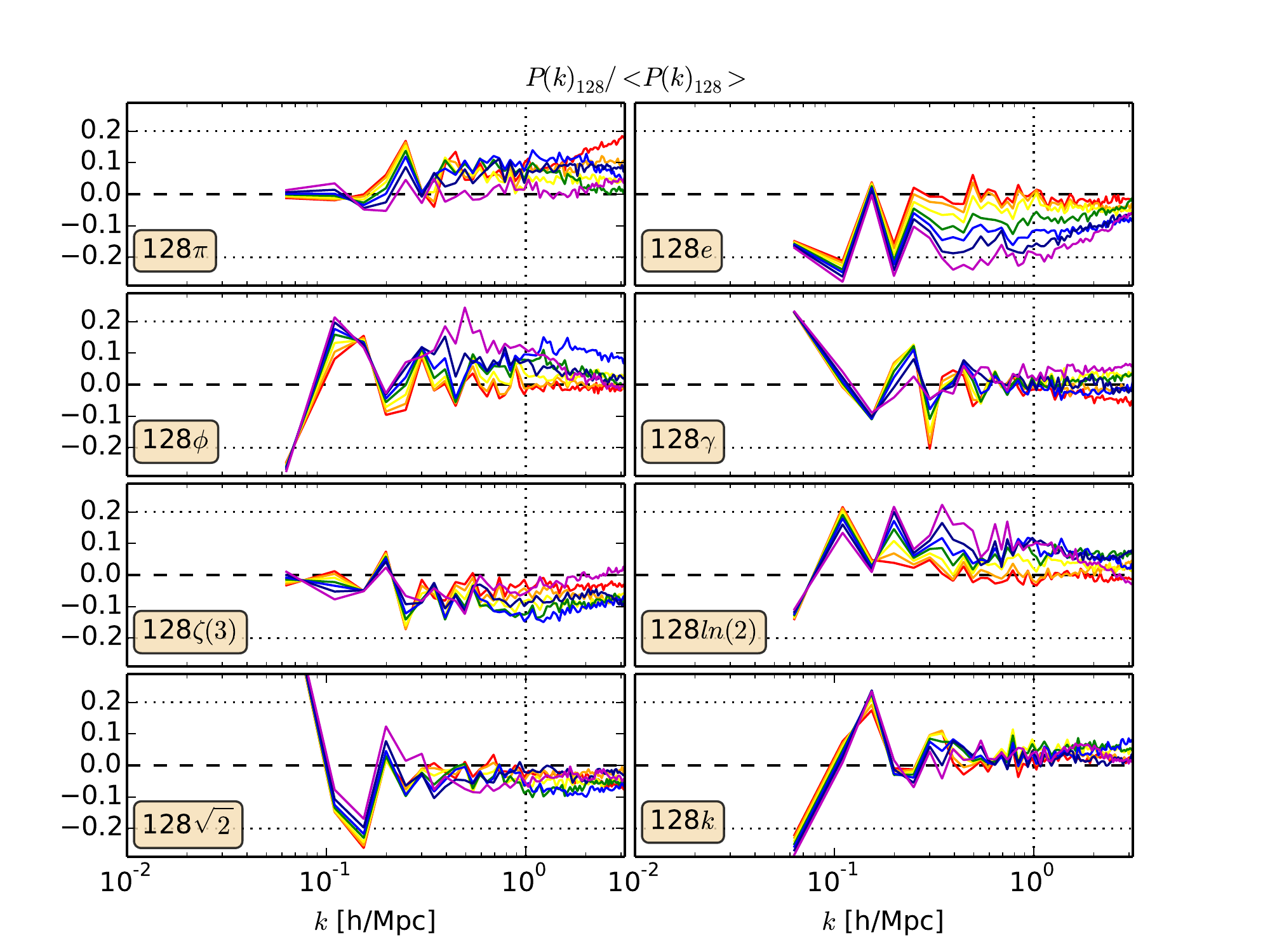}
\caption{Simulations spectra of different realizations (dubbed:
  $\pi,~\phi,$$~\zeta(3),$$~\sqrt{2},~e,$$~\gamma,~ln(2),k$) in a 
  128 box, vs.~their averages at 8 different redshifts ($z=3.00$
  (red), $2.20$(orange), $1.5$(yellow), $1.00$(green), $0.7$(blue), 
  $0.35$(indigo), $0.00$(violet)).
\label{all128} }
\end{figure}
\begin{figure}
\epsscale{.6}
\plotone{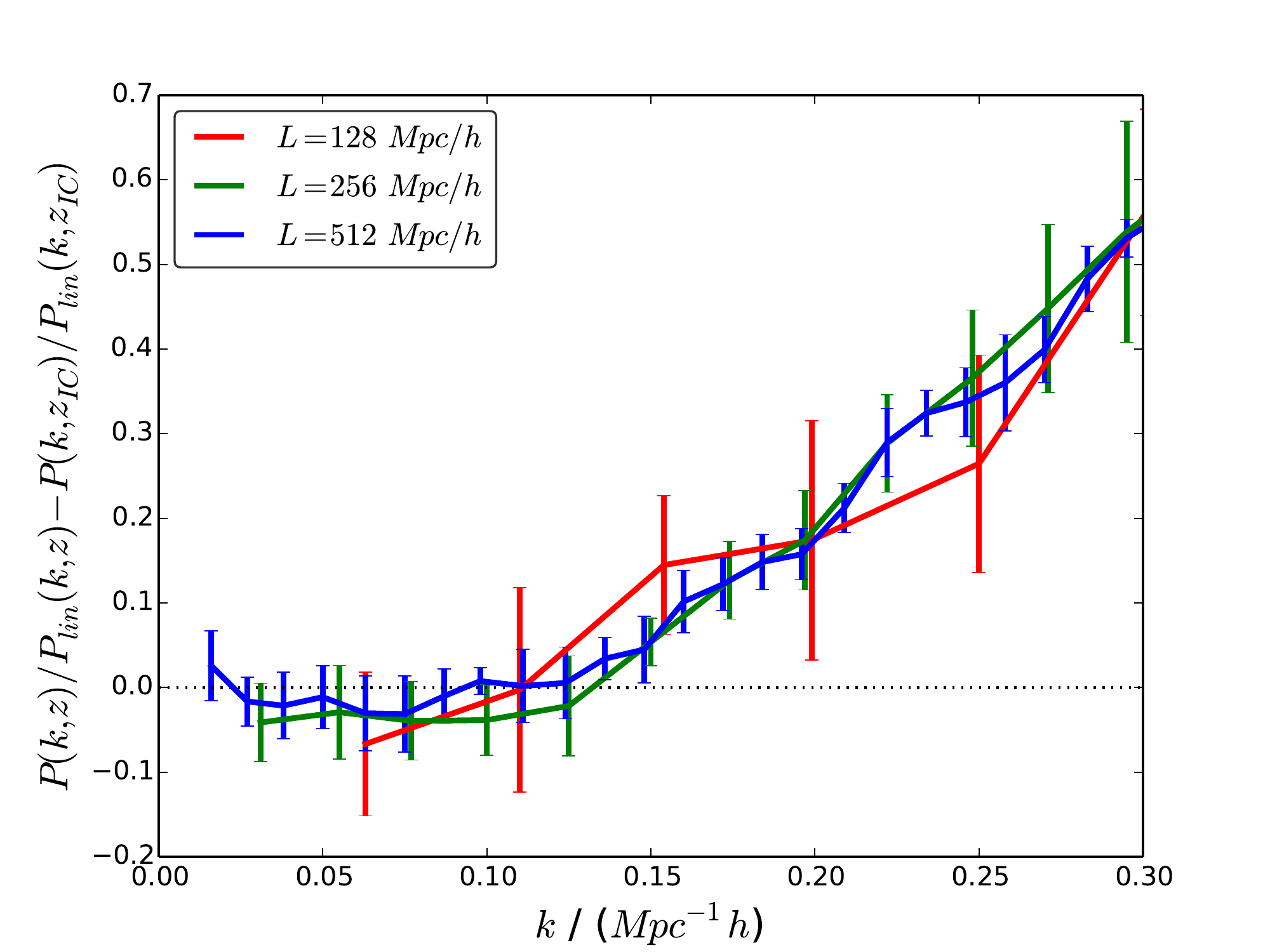}
\caption{Average of spectra, after subtracting initial conditions
from each of them, compared with linear growth.
\label{subtract} }
\end{figure}

In what follows we shall mostly report results obtained from
simulations in 128 and 512$\, h^{-1}$Mpc boxes, with $n$ 1024 and
4096, respectively. Some results from 256$\, h^{-1}$Mpc box with
$n=2048$ will only be cited in the final Section.

At low $k$ spectral values obtained form simulations still exhibit a
significant dependence from the seed yielding Initial Conditions. This
is clearly visible in Figure \ref{all128}, referring to the 128$\,
h^{-1}$Mpc box; here we make use of the spectra arising from the
different seeds, dubbed: $\pi,~ \phi,$$~ \zeta(3),$$~ \sqrt{2},~ e,$$~
\gamma,~ ln(2),~k$; these symbols are preceded by the related box side
in $\, h^{-1}$Mpc. For instance, for the box with 128 (512)$\,
h^{-1}$Mpc aside, the simulations 128$e$ and 128$\gamma$
(512$\zeta(3)$ and 512$\sqrt{2}$) will be used. (When needed for
graphic reasons these names are shortened, e.g., by omitting the box
side or even by replacing $\zeta(3)$ or $\sqrt{2}$ by
$\zeta$~or~$\sqrt{} $.)  

In Figure \ref{all128} we considered $P(k,z)$ up to $k \simeq 0.32\,
h\, {\rm Mpc}^{-1}$ i.e., approximately, up to where spectral
discreteness is visible on a logarithmic plot, when plotting
$P(k)/\langle P(k) \rangle$; the denominator being an average among
different realization spectra.  This allows then us to appreciate that
deviations from such average, due to the seed, are persistent at all
subsequent $z$'s, starting from initial conditions.

Little use of $\langle P(k) \rangle$ will however be made in spectral
analysis. Instead of using it, we shall rather refer to the actual
realization closest to it. Then sample variance is estimated by
comparing its spectra with those of the realization most distant from
$\langle P(k) \rangle$ itself. According to the discussion in the
previous Section, we shall neglect the discrepancy between a full
ensemble average and the 8--average considered, as well as the
discrepancies between the spectrum closest to $\langle P(k) \rangle$
and $\langle P(k) \rangle$ itself. These neglects will however allow
us to treat the two simulations on the same footing.

The relevance of this approximation can be better appreciated by
averaging among simulation octets, and comparing spectra from
different box sizes, as is done in Figure \ref{subtract}. More
precisely, in this Figure we consider the ratio $P(k,z)/P_{lin}(k,z)$,
between the spectrum of simulations and the linear spectrum; we
subtract from it the same ratio at $z_{IC}$, where IC were set;
finally we average among the 8 realizations obtained from different
seeds, for the 3 box sizes. This operation aims at minimizing the
discreteness jumps which, as seen in the previous Figure \ref{all128}
persist through redshifts, further smoothing the result by averaging
among seeds. Error bars yield the sample variance at 1--$\sigma$.

The Figure allows us to confirm that the DC term we omitted in Initial
Conditions implies a correction to variance substantially smaller than
the expected discrepancy between estimated and actual sample
variance. In fact, in the low--$k$ almost--linear range, although some
lack of power of smaller boxes seems appreciable by eye, the sample
variance at all $k$'s comfortably includes the zero point;
furthermore, as soon as we enter the non--linear regime, the Figure
shows the lack of any apparent trend indicating smaller box spectra to
have less power than larger ones. On the contrary, in the initial
mildly non--linear regime, the greatest fluctuation amplitudes are
those obtained from the smallest box while, along the whole scale
range shown, the greatest box average spectra never are the top ones
(for an example of greater statistics making the DC term significant
see, e.g., the plot in Figure 6 of \cite{Heitmann et al. 2010},
analogous to Figure \ref{subtract}).

Let us now turn to the main aim of this work, detecting the impact of
variance on shear spectra $C_{ij}(\ell)$. To do so we must treat each
realization separately, ignoring other ones, by devising a recipe
enabling us to use the spectra $P(k,z)$, obtained from each single
realization, separately.

Each single spectrum, as derived from simulations, exhibits problems
at small $k$, as we just appreciated, as well as large $k$ problems.
The latter ones may be due either to lack of resolution --due to mass
discreteness-- or to numerical noise. In fact, the signal due to
initial lattice can cover the $P(k)$ signal at high $z$'s if, as is
usual, initial conditions are given on a grid rather than on a glass.
The recipe to deal with this problems will be detailed in the next
Subsection.

Large $k$ problems, however, are not so essential as those found at
low--$k$ which, in turn, are quite different from those faced to build
an {\it average} non--linear spectrum, at a given approximation level,
as done, e.g., within the Coyote simulation suite \citep{Heitmann et
  al. 2010}. In fact, here we need to preserve the peculiarities
deriving from each specific model realization in the assigned box,
which are the essential feature whose impact on shear spectra we wish
to gauge. In turn, at low $k$, each spectral point being derived by
averaging over quite a limited number of realizations allowed inside
the box size, significant jumps upwards and downwards, as those shown
in Figure \ref{all128}, are unavoidable.

\begin{figure}
\epsscale{.7}
\plotone{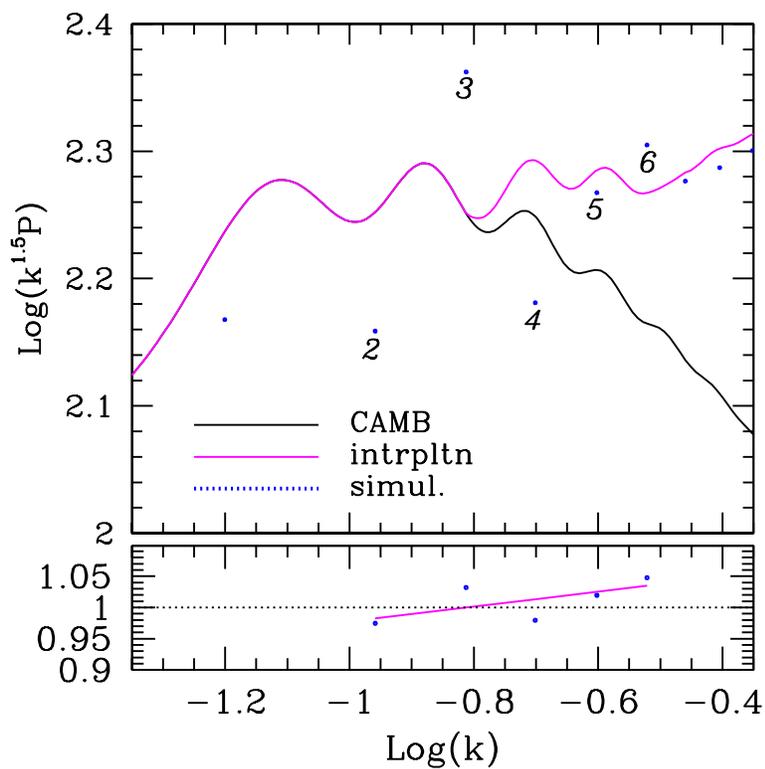}
\vskip -2.truecm
\caption{Low--$k$ interpolation between CAMB and numerical spectra for
  a box side $L=128\, h^{-1}$Mpc at $z=0~.$ The interpolatory curve in
  the upper frame is obtained by displacing the CAMB spectrum upwards,
  as soon as the linear interpolation of the ratio $\cal R$$(k_i)$ (see
  text) exceeds unity. In the bottom plot $\cal R$$(k_i)$ values and
  its linear interpolation are showed. All units are suitable powers
  of $h^{-1}$Mpc.\label{FC} }
\end{figure}
\begin{figure}
\epsscale{.8}
\plotone{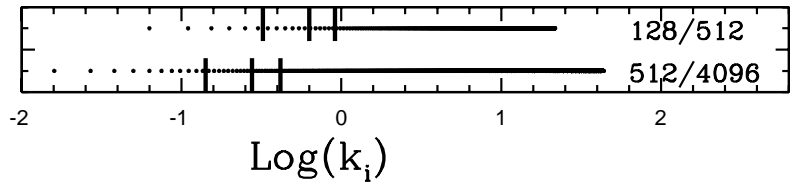}
\vskip -2.truecm
\caption{Points indicate the $k_i$ for which $P_S$ spectra are
  obtainable, when using a 128 or 512 $h^{-1}$Mpc box. Vertical lines
  indicate the limits of the 3 intervals where $P_S$ is interpolated
  with $P_C$, according to procedures outlined in the text. For the
  128 (512) $h^{-1}$Mpc box the number of points in the two intervals
  is 6 (11).
\label{tr}}
\end{figure}

The recipe to be used at low $k$ will also be discussed here below and
detailed in the next Subsections. Let us soon outline that no use of
perturbative expressions of mild spectral non--linearity will be made.

As a matter of fact, a large deal of work on this subject has been
carried on along more than three decades, since Peebles (1980) book
and Bernadeu et al (2002) review. Recent work
\citep{Crocce,Anselmi1, Anselmi2}
shows that corrections to the linear spectrum approaching $1\, \%$ may
be present up to $k \sim 0.003\, h$Mpc$^{-1}$, i.e. over scales $\sim
4 $ times greater than those inspected by our greatest simulation
box. More significantly, they extend perturbative techniques to $k$
values well beyond the BAO range, where non--linearity apparently
dominates.

However, apart of the difficulty to extract from these results a
handable parametric expression, such summation techniques apply to
full ensemble averages. Possibly, their use could be effective to
improve results obtained from averaging among a significant number of
simulations. On the contrary, they are unlikely to apply to the
results of single simulations.

Accordingly, we shall keep on the phenomenological side, and make use
of expression bearing a purely analytical significance. Incidentally,
when using our 512$\, h^{-1}$Mpc box, they univocally find that
convergence on the linear results is attained for $\log(k/h{\rm
  Mpc})^{-1}) \simeq -1.2,$ in agreement with \cite{Matsubara 2008}
and \cite{Carlson et al 2009}.

The basic steps of our technique are as follows: we first interpolate
the linear spectra to obtain their values $P_C(k_i)$ at the very
$k_i$'s where the simulation spectrum $P_S(k_i)$ is calculated. We
then fit the ratio
\begin{equation}
{\cal R}(k_i) = \log[P_S(k_i)]/\log[P_C(k_i)]
\end{equation}
with a curve growing linearly with $k$, for the first few points,
allowing for a more detailed correction after a few of them. We do so
we neglect the first point $k_1$ and refer to different numbers of
$k_i$'s for the different box sizes, although selecting them through
fixed rules. In order to be more precise, let us now distinguish
between different box sizes.
 
\subsection{Simulations in 128$\, h^{-1}$Mpc box}
The hardest case is the 128$^3$ particle simulations, for which we
provide some more details. In this case we start from the $k_i$ values
with $i$ from 2 to 6 (five values) and determine the $P$ and $Q$
coefficient minimizing the expression
\begin{equation}
diff = \sum_i \left[(P + Qk_i)/{\cal R}(k_i) - 1 \right]^2~,
\end{equation}
adding a specific condition, soon specified here below. We then {\it
  fit} the simulation spectrum with the expression
\begin{equation}
\log[P_{fit}(k)] = \log[P_C(k)] \times (P + Qk) 
\end{equation}
for all $k$ values where it exceeds $\log[P_C]$. The specific
condition outlined here above, is that the spectrum must however meet
the linear spectrum at any $k$ smaller than a suitable $ \bar k$. The
point is not only that non--linear effects are surely (almost) absent
below such suitable $k$ value, but that simulations cannot provide
information on possible (residual) non--linearities for length scales
$2\pi/k$ too close to the box side. For our box side $L_{box} = 128\,
h^{-1}$Mpc, we then take $\bar k = 0.1\, h\, $Mpc$^{-1}$ (yielding a
length scale $\sim L_{box}/2)~.$
\begin{figure}
\epsscale{.9}
\vskip -2.5truecm
\plotone{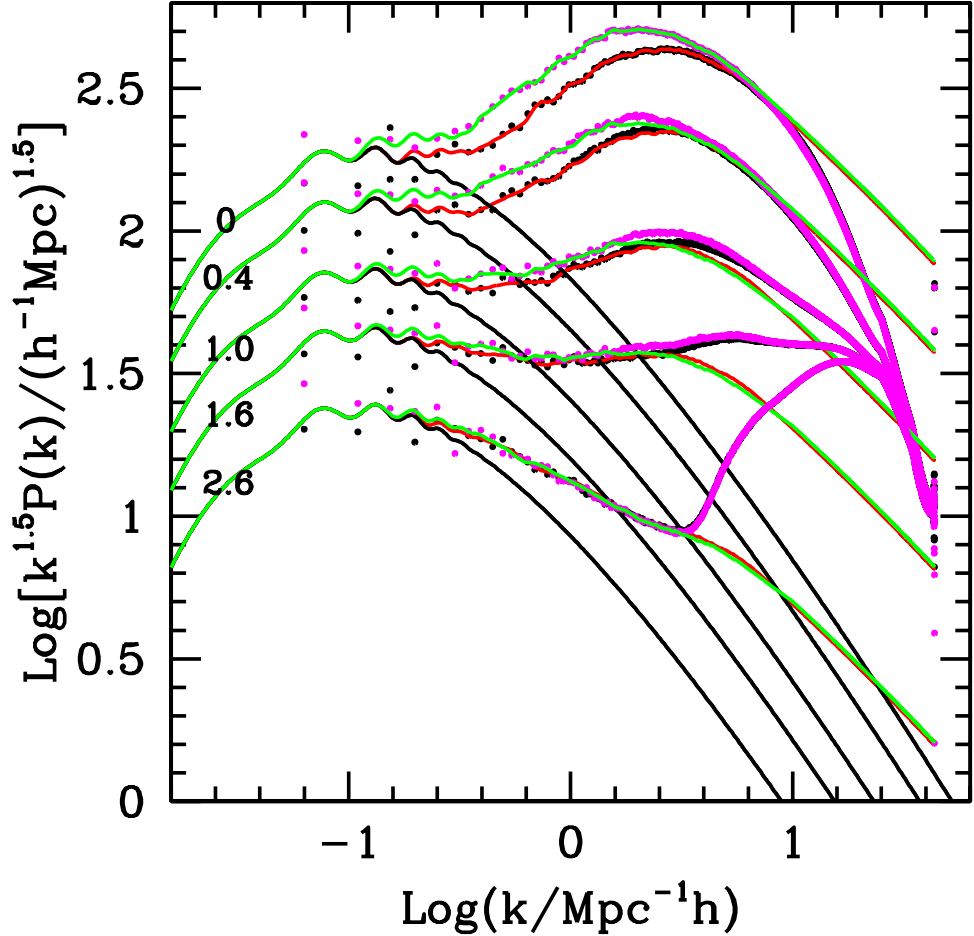}
\vskip -4.truecm
\caption{Spectra obtained from simulations in a box with $L=128\,
  h^{-1}$Mpc aside, interpolated with linear CAMB outputs. The main
  feature shown in this figure is the significant discrepancy between
  the two seeds, dubbed $128\gamma$ (closest to average) and $128e$
  (farthest from average and smaller from it). Magenta and black
  points, respectively, yield the resulting simulation outputs. From
  CAMB and them we derive the green and red spectra (also
  respectively), as explained in the text.
  \label{z4.128} }
\end{figure}
\begin{figure}
\epsscale{.9}
\vskip -2.5truecm
\plotone{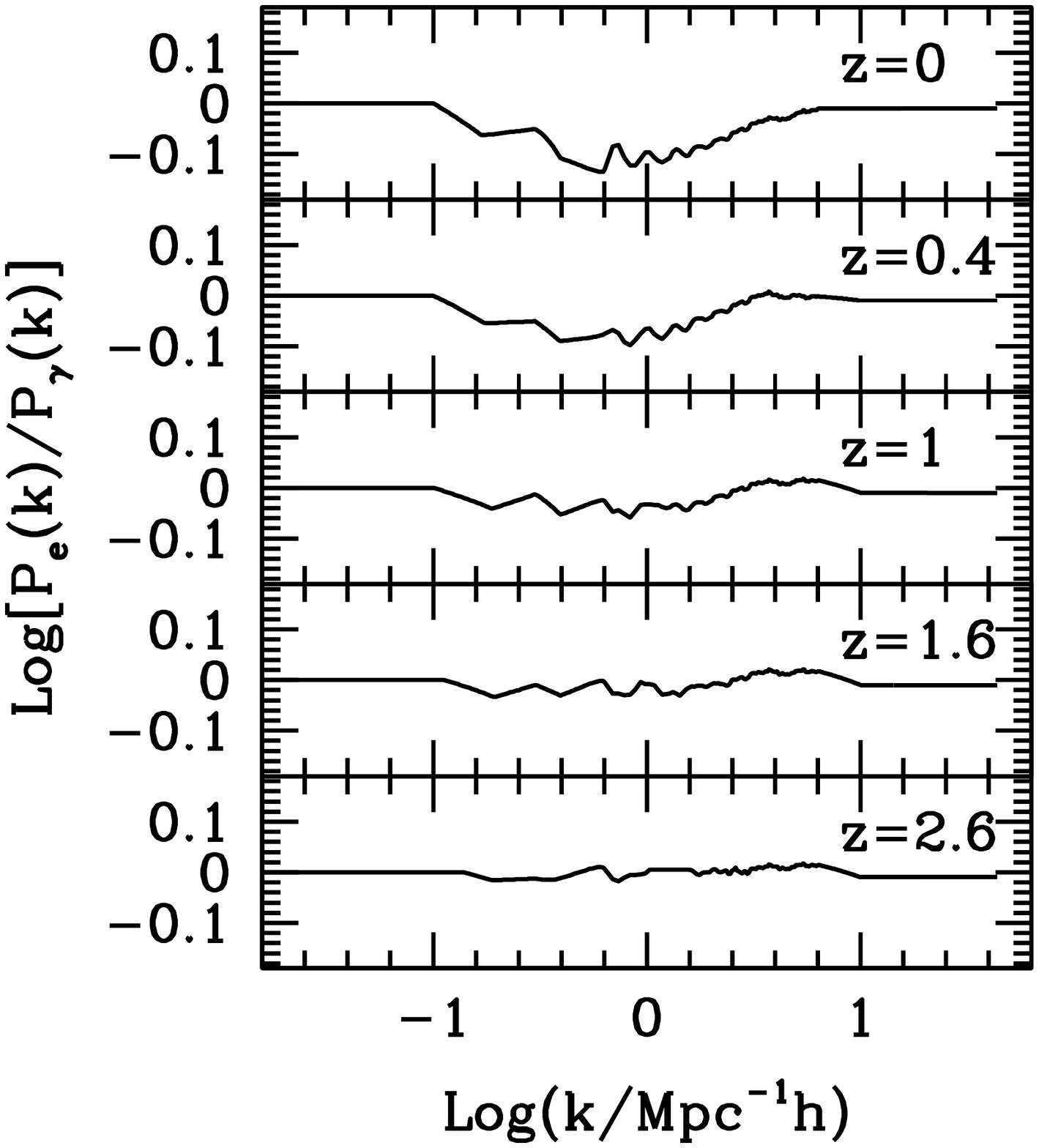}
\vskip -4.truecm
\caption{Ratio between the spectra deriving from the seeds $128e$ \&
  $128\gamma$, at the redshifts indicated in the frames. The top
  discrepancy shown, at $z=0 $ and $k/(h\, {\rm Mpc}^{-1}) \sim 0.65$,
  is $\simeq 37.4\, \%$.
  \label{Pratio} }
\end{figure}

Fig.~\ref{FC} gives more details and shows the results of this
operation. In the upper panel we show the overall interpolation
(magenta curve). In the lower panel we show the ratios ${\cal R}(k_i)$
and their linear interpolation $P + Qk$, to be used where it exceeds
unity (the magenta line). This technique is meant to preserve the BAO
structure outlined by the linear algorithm, just suitably shifting it
upwards, according to the requirements coming from the first $k_i$
values in the simulation.

We then consider 6 more points and allow for a correction to $P+Qk$ by
a term $R (k-k_6)^\alpha$, $R$ and $\alpha$ being again determined
through a l.s. fit. This is meant to allow a progressive rise of the
spectral steepness, following the gradual incoming on non linear
dynamics. Accordingly, this approach must be gradually modified at
higher $z$, when non linear dynamics does not yet affect smaller
$k_i$'s. As a matter of fact, at fairly large $z$ values, not only the
$P$ and $Q$ coefficients may turn out to be quite small, but there may
be no need of power law corrections. Then, if $P_{fit}(k)$ exceeds
$P_S(k)$ in two --or more-- $k_i$ points ($7 \leq i \leq 12$), we shall
deal with these points as we do with those above the 12--th, and is
described below.

In Figure \ref{tr} we show the limits on the $k$ axis of the 
two above intervals, both in this case and for the forthcoming
512$\, h^{-1}$Mpc box.

Starting from $k_n$ with $n \geq 13$, we then perform a
Savitski--Golay (SG) interpolation\footnote{Numerical Recipes,
  Cambridge U. Press 1986, 1992; Sec.~14.8}. More precisely, we
consider $2p$ values $k_{n\pm j}$ ($j=1,..,p$), plus $k_n$ itself, and
interpolate them to obtain $2p+1$ equispaced points on a $\log k$
scale, with extremes in $k_{n\pm p}$. In this way we work out a
spectral value for $k'_{n} = 10^{[\log(k_{n-p})+\log(k_{n+p})]/2}$, a
point quite close to $k_n$ but not coinciding with it. The values of
$p$ go from 4 to 8, suitably increasing towards greater $k_n$ values.

Before passing to briefly describing the large--$k$ treatment, let us
still outline that the same treatments are reserved to all seeds. In
particular, when $P_{fit}(k)$ exceeds $P_S(k)$ for almost 2 $k_i$
points ($7 \leq i \leq 12$) in one simulation, we start operating a SG
interpolation for all of them.

As far as large--$k$ and low $z$ are concerned, resolution problems
begin to damp the spectrum at $k/h\, {\rm Mpc}^{-1} \sim 3\, $. On the
contrary, at large $z$, numerical noise cancels spectral features;
e.g., at $z = 2.8$, this occurs at $k \sim 5\, .$

The problems we meet, therefore, concern a range where physical signal
cannot be predicted by using gravitation only. Accordingly, our
treatment is meant just to test sample variance in shear spectra; as
any previous weak lensing treatment, based on N--body simulations
only, the values found for the shear spectra start to be biased as
soon as $\ell >\sim 500$ (see also Figure 12, herebelow).

Accordingly, we implement an algorithm detecting two possible spectral
anomalies: (i) Non linear spectra decreasing more rapidly than CAMB
linear spectra $\times k^\alpha$, when $k$ increases (the results are
obtained with $\alpha=0.8$, in rough agreement with a generic trend
visible also in \citep{halofit}). (ii) Spectra exhibiting an
increasing steepness, well after non--linearity has onset.

In both cases, our algorithm gradually replaces numerical spectra with
curves parallel to the linear CAMB spectrum $\times k^\alpha$. Here
again, we abandon the numerical spectrum at the same $k$ values for
all seeds.

In Fig.~\ref{z4.128} we show the results of these operations for 2
simulations in boxes of side $L = 128\, h^{-1}$Mpc. These simulations
are dubbed $128\gamma$ and $128e$ in Figure \ref{all128}, and are
those closest and farthest from average, respectively.

The size of the discrepancy between these simulations can be further
appreciated in Figure \ref{Pratio}, where we plot the ratio between
the spectra obtained from the two seeds, and its low $z$ evolution.
Its top value, estimated by considering averages among $N$ points
($N=6$) amounts to 37.4$\, \%~.$


\subsection{Simulations in 512$\, h^{-1}$Mpc box}
\begin{figure}
\epsscale{.9}
\vskip -2.truecm
\plotone{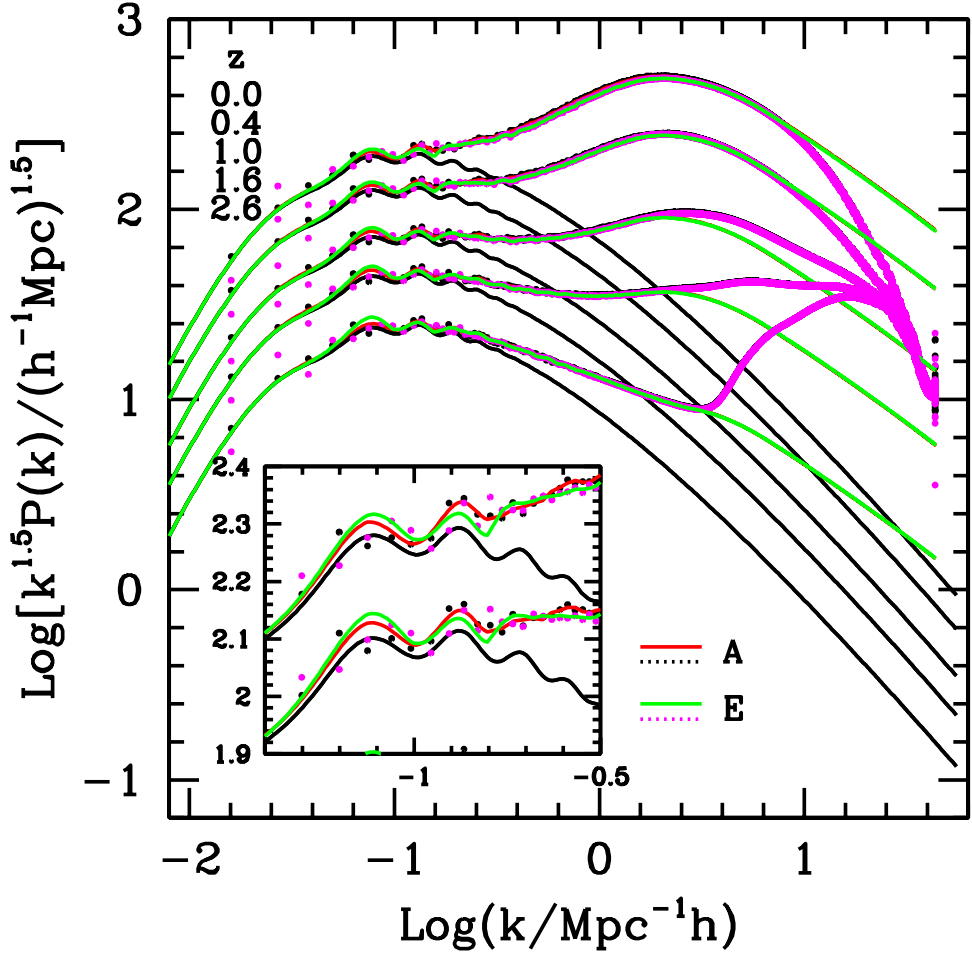}
\vskip -4.truecm
\caption{Spectra obtained from simulations in a box with $L=512\,
  h^{-1}$Mpc aside, interpolated with linear CAMB outputs. The seeds
  closest and farthest from (mostly greater than) average are those
  dubbed $512\sqrt{2}$ and $512\zeta(3)$. Here again, spectra in black
  are CAMB outputs.  Black and magenta points yield actual simulation
  outputs. From CAMB and them we derive the red and green spectra
  (respectively), as explained in the text. Notice also the
  significant spectral discrepancy occuring at small $k$ values,
  magnified in the inner box; for a significant $k$--interval, its
  sign is opposite in respect to the main trend, as the very
  ``points'' of spectrum A, which lay above the ``points'' of spectrum
  E.
 \label{z4.512}}
\end{figure}
\begin{figure}
\epsscale{.9}
\vskip -2.5truecm
\plotone{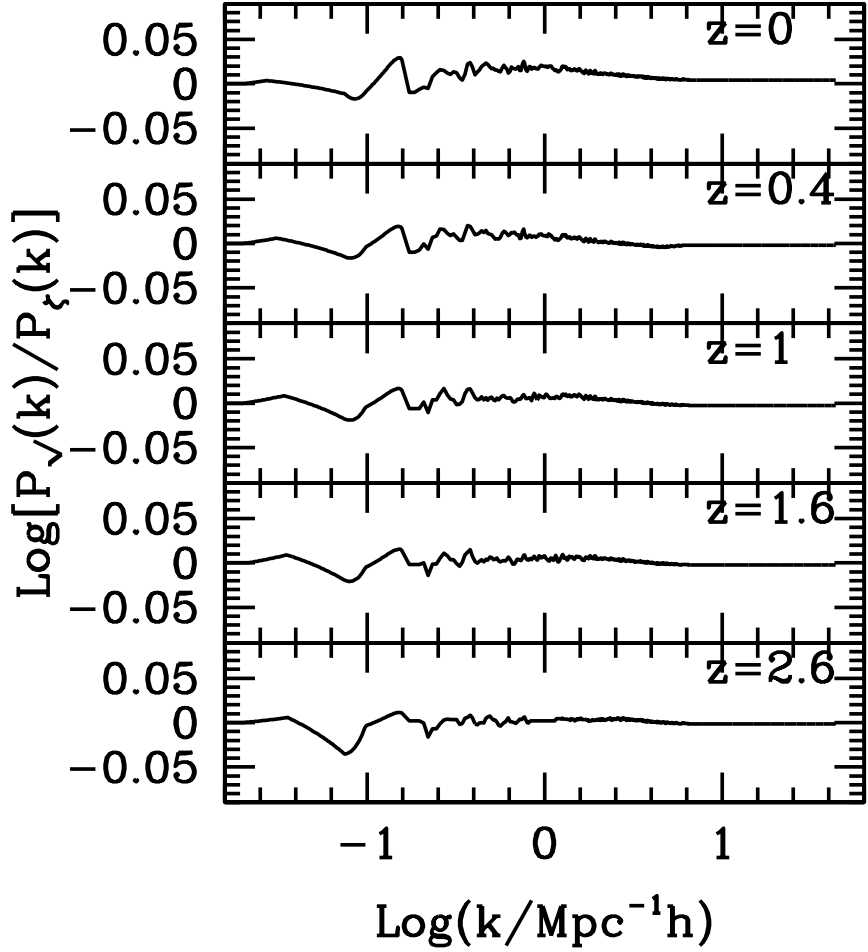}
\vskip -4.truecm
\caption{Ratio between the spectra from the seeds $512\sqrt{2}$ \&
  $512\zeta(3)$, at the redshifts indicated in the frames. Notice the
  ordinate scale expanded in respect to the analogous Figure 7. Jumps
  at low $k$ have alternate signs and arise in the linear/non--linear
  connection region where the box still contains a limited number of
  realizations.  As explained in the text, they have a lesser impact
  than the steady excess of $P_{\sqrt{2}}$ on $P_{\zeta(3)}$ at
  greater $k$ values.
  \label{Pratio1} }
\end{figure}

The same tecnique used for the smaller box is now extended to this
wider box, to allow a linear/non--linear connection. In Figure
\ref{tr} we compare the range of $k$ values used in the two cases.
The number of ``points'' used in the low--$k$ intervals is now 11.
Tries performed with close numbers of points (e.g. 10 or 12) yield
greater fitting residuals.

Figure \ref{z4.512} is then analogous to Fig.~\ref{z4.128}, although
spectral ``points'' here extend down to smaller $k$ values, just
marginally non--linear. These points are indeed obtained by averaging
over a limited number or realization, so that sample variance is large
and partially hides non--linearity itself. In the inner box this scale
range is magnified, also showing the smoothness of the
linear/non--linear interpolated red and green curves, however
crossing a few times.

The overall situation is more clearly outlined by Figure
\ref{Pratio1}, analogous to Figure \ref{Pratio}; notice however the
reduced range of the ordinates. Here we clearly distinguish the
effects due to the limited number of realizations at low $k$, which
persist through all $z$ without an appreciable amplitude growth, from
the actual sample variance between the two seeds, in the $k$ interval
(-0.6 -- +0.7), steadily growing because of non--linearity: at $z=2.6$
the spectral ratio is still $\sim 1\, \%$; at $z=0$ it finally reaches
$5.32\, \%$ (evaluated with the same criterion as from Figure
\ref{Pratio}).

\section{Tomographic windows}
Let us now investigate how these variance effects transfer from
fluctuation to shear spectra. To do so, let us build the needed
tomographic windows, by starting from the expression of the background
metric reading
\begin{equation}
ds^2 = a^2(\tau)[d\tau^2 - d\lambda^2]~,
\label{metric}
\end{equation}
$d\lambda$ being the comoving space element, while $\tau$ is the {\it
  conformal }time; both are measured in $h^{-1}$Mpc. Let then $a(\tau)
= 1/(1+z)$ be the scale factor, $z$ being the redshift. Let us further
define the {\it conformal time delay} at redshift $z$
\begin{equation}
u(z) = \tau_0-\tau(z)~,
\label{uz}
\end{equation}
$\tau_0$ being the present conformal time, and the inverse function
$z(u)$.
\begin{figure}
\epsscale{.7}
\vskip -6.truecm
\plotone{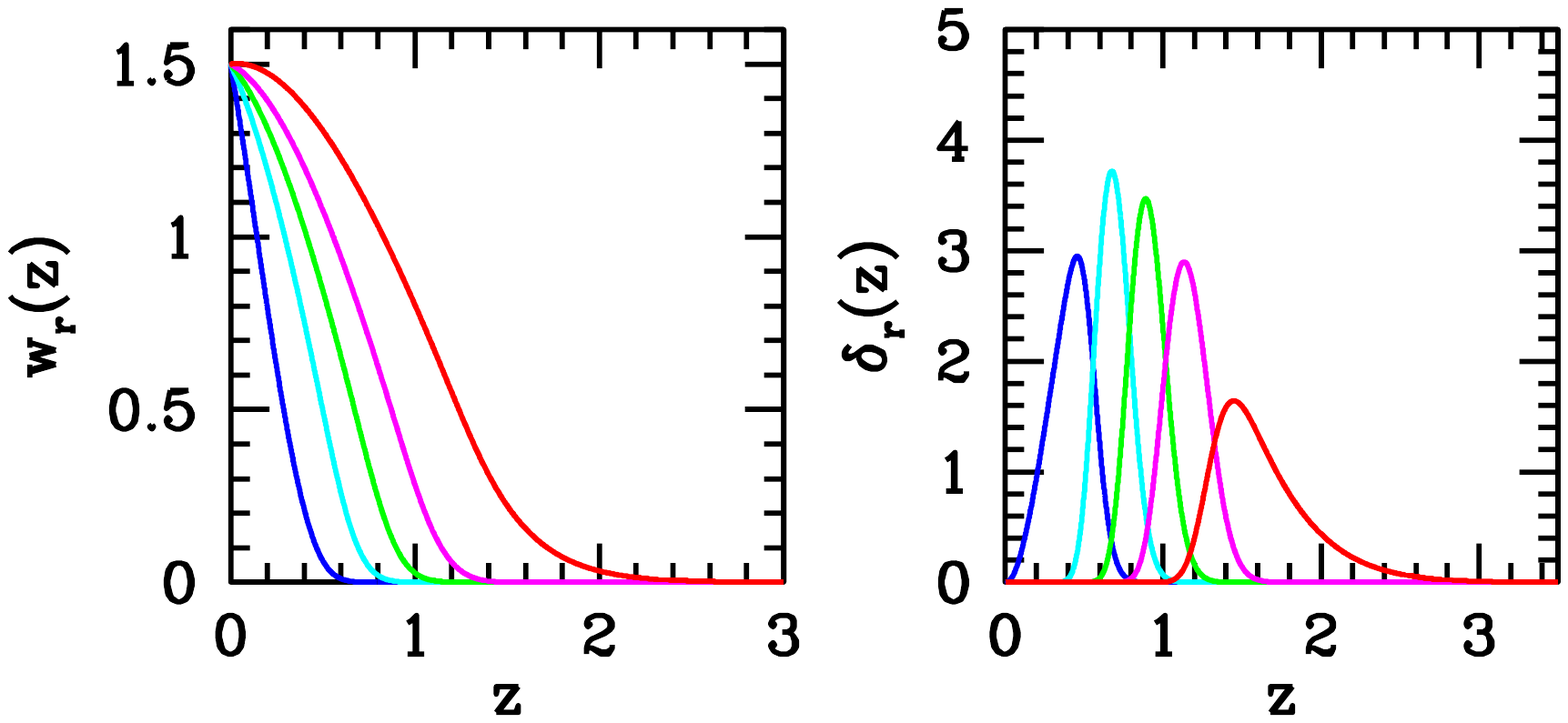}
\vskip -9.truecm
\plotone{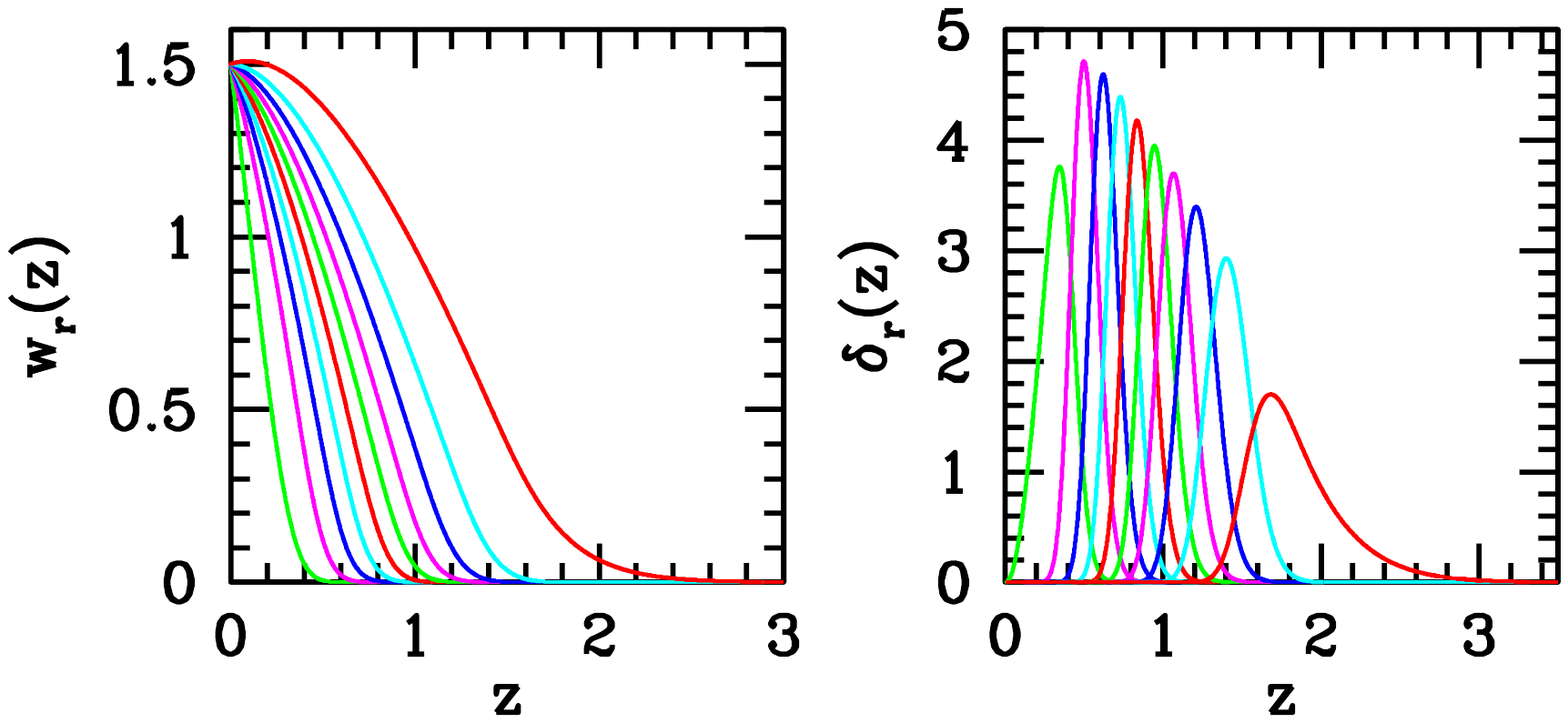}
\vskip -4.truecm
\caption{Expected distributions on redshift of galaxies belonging
to different bands (l.h.s.) and related filters (r.h.s.). Upper
(lower) panel refers to the 5 (10) band case.\label{dW}}
\end{figure}

Galaxies observed in a unit solid angle are then assumed to have a
redshift distribution
\begin{equation}
n(z) = {d^2 N \over d\Omega\, dz} = {\cal C} \times \bigg({z \over z_0
}\bigg)^A \exp\bigg[- \left( z \over z_0 \right)^B \bigg]
\end{equation}
with  $A=2$, $B=1.5$ so that
\begin{equation}
{\cal C} = {B \over \left[z_0 \Gamma \left( A+1 \over B
\right) \right]}  = {1.5 \over z_0}
\label{nz}
\end{equation}
(here $z_0 = z_{m} /1.412$ while the median redshift $z_m = 0.9~$, in
agreement with {\sc Euclid} specifications).

These galaxies will be shared in $N=5$ or 10 redshift bands, whose
limits $z_r$ are selected so that they contain equal galaxy numbers.
For large galaxy sets, photometric redshifts only are expected to be
available and, to evaluate the expected distribution on redshift for
the $r$--th band galaxies, we apply the filters
\begin{equation}
 \Pi_r (z) = \int_{z_{r}}^{z_{r+1}} dz' ~ {\exp \left[-{(z - z')^2
      \over 2\, \sigma^2 (z)} \right] \over \sqrt{2 \pi}~ \sigma(z)} ~=~
{1 \over 2}
\left[
{\rm Erf}\left(z_{r+1}-z \over \sqrt{2}\sigma(z)
\right)-{\rm Erf}\left[z_{r}-z \over \sqrt{2}\sigma(z)\right]
\right]
\end{equation}
to $n(z)$. In this way we obtain the distributions
\begin{equation}
D_r(z) = n(z) \Pi_r(z)
\label{diz}
\end{equation}
whose integrals are $\simeq 1/N$. In this work we shall take
$\sigma(z) = 0.05~(1+z)$, coherently with {\sc Euclid} expectations
(\cite{Amendola et al. 2013}, see also \cite{Casarini et al 2011});
the distributions $D_r$, when normalized to unity, are then dubbed
$\delta_r$~.
They are used to define the {\it window functions}
\begin{equation}
\label{WI}
w_r (u) = {3 \over 2} [1+z(u)] \int_u^\infty du'~\delta_r(u')~ {u'-u
  \over u' }~,
\end{equation}
also shown in Figure \ref{dW}.

\section{Shear spectra}
Shear spectra are then related to the power spectra $P_\delta(k,u)
\equiv P[k,z(u)]$, through the relation
\begin{equation}
C_{ij}(\ell) = (H_{0}^{2} \Omega_{0m})^2 \int_{0}^{\tau_0 } du~ w_{i}(u)
\, w_{j}(u)~ P_\delta(\ell/u,u)~.
\label{pijl}
\end{equation}
\begin{figure}
\epsscale{.7}
\vskip -4.truecm
\plotone{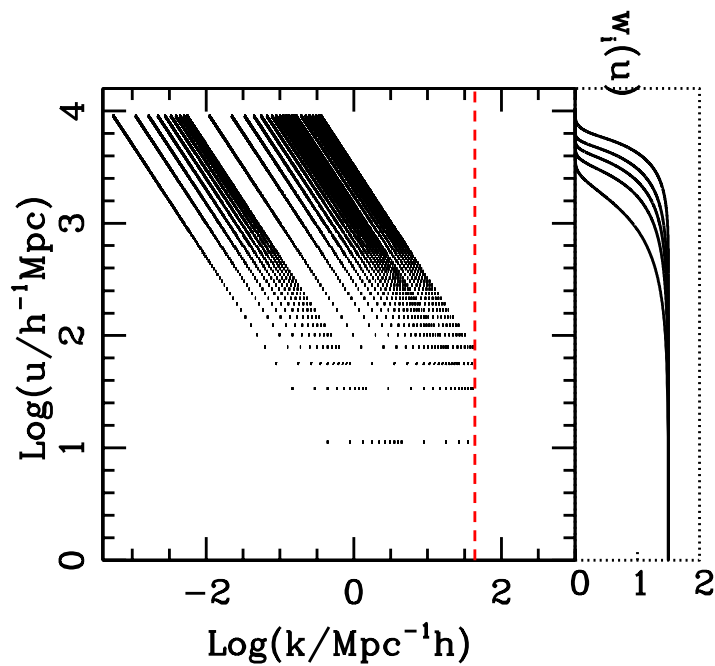}
\vskip -3.truecm
\vskip -.5truecm
\caption{Integration patterns for a number of $\ell$ values between 5
  and 3300. Points indicated are those used in the case of Riemann
  integration with 400 points (yielding results visually
  indistinguishable from those with more points). The red dashed line
  is the limit beyond which (just a few) integrand values are to be
  obtained by interpolation. In the right frame we show the patterns
  of window functions $w_i(u)$, in the case of 5 bands.}
 \label{klnew} 
\end{figure}
Integrals are performed by using a modified Riemann algorithm with
10000 equispaced integration points $u_\alpha$ up to $u_{top} \simeq
9000$ (instead of $\tau_0$).  Integration results are visually
indistinguishable from those obtained with just 400 points and, for
this case, in Figure \ref{klnew} we show the points selected on the
$\log k$--$\log u$ plane, for a subsample of $\ell $ values, lying
along tilted straight lines.

Integration therefore requires interpolation of $w_j(u)$ at $u_\alpha$
and of $P_\delta$, first along $k$ and then again along $u$.
A number of cases is then considered: for 5 and 10 tomographic bands;
for simulations in 128(, 256) and 512$\, h^{-1}$Mpc boxes, with
resolution pushed up to 1024(,~2048) and 4096 points, respectively;
for those 2 seeds, selected for each box size, for being closest and
farthest from average.

\subsection{Shear spectra from simulations in the 128$\, h^{-1}$Mpc box }
\begin{figure}
\epsscale{.9}
\vskip -1.3truecm
\plotone{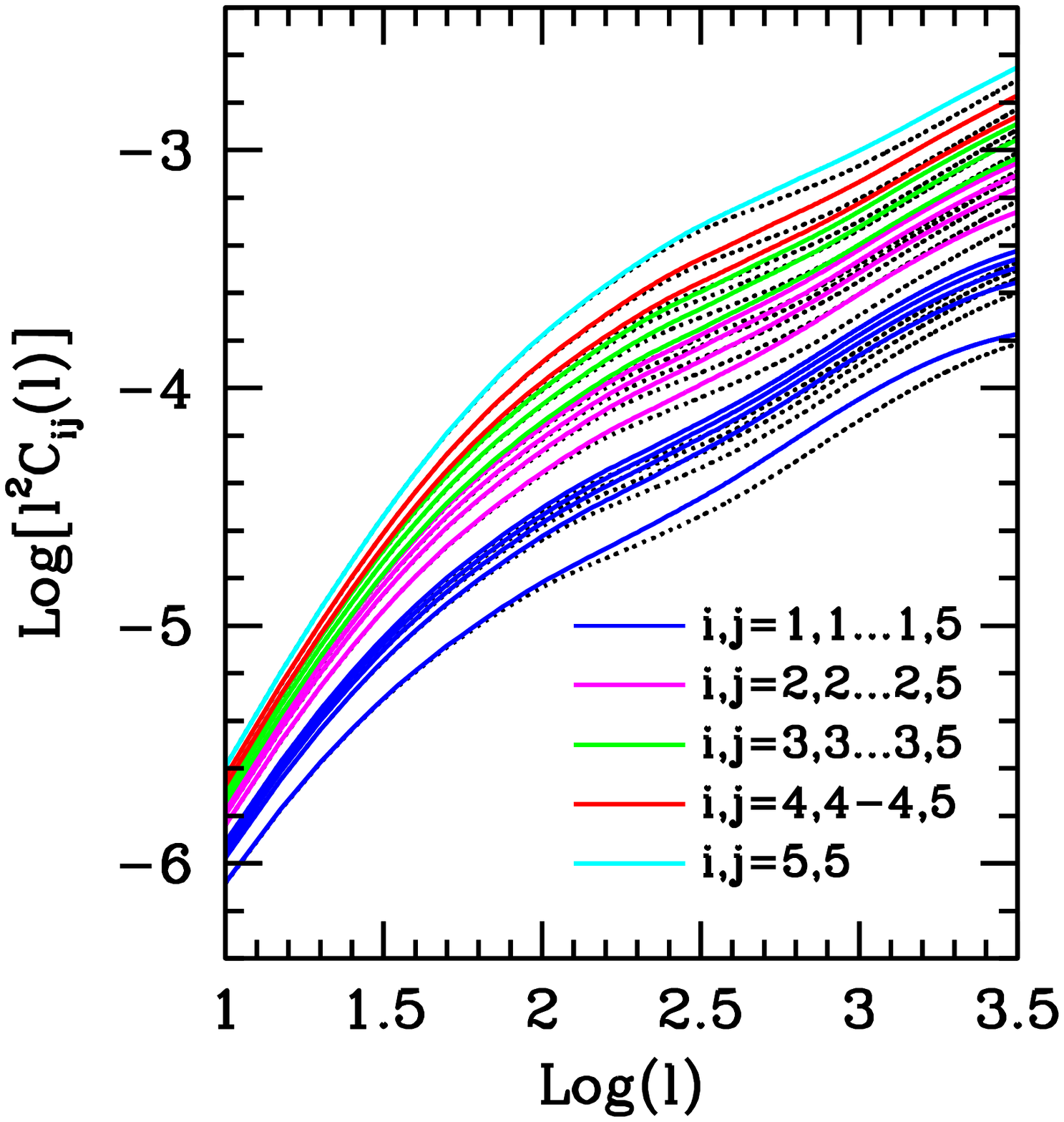}
\vskip -1.9truecm
\caption{Solid lines yield the shear spectra $C_{ij}(\ell)$ for the
  128$\gamma$ seed, in the case of a 5 band tomography. Colors account
  for $_{i,j}$ indices. Almost overlapped to them we also plot the
  spectra obtained from the $128e$ seed (black dotted curves), so
  outlining the significant discrepances between the two seeds.  }
 \label{128-5} 
\end{figure}

\begin{figure}
\epsscale{.9}
\vskip -1.3truecm
\plotone{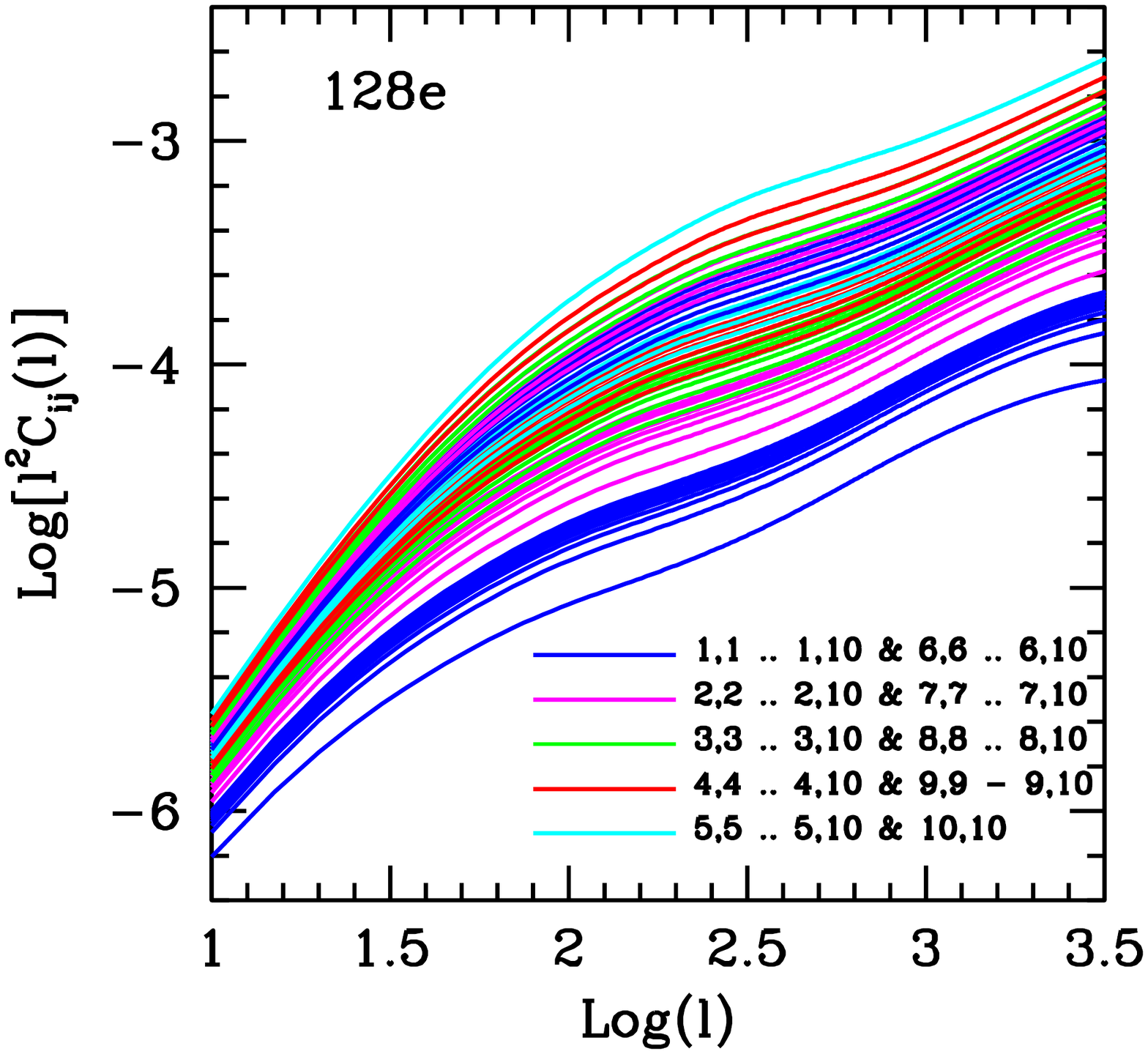}
\vskip -1.9truecm
\caption{Shear spectra $C_{ij}(\ell)$ for the 128$e$ seed, in the case
  of a 10 band tomography.  }
 \label{128-10} 
\end{figure}

In Figure \ref{128-5} we plot the shear spectra obtained from the
fluctuation spectra arising from the seed $128\gamma$, in the case of
a 5 band tomography. Aside of them we also show (black dotted curves)
the spectra obtainable from $128e$. In Figure \ref{128-10} we then
show the spectra for a 10 band tomography obtained from $128e$.

\begin{figure}
\epsscale{.9}
\plotone{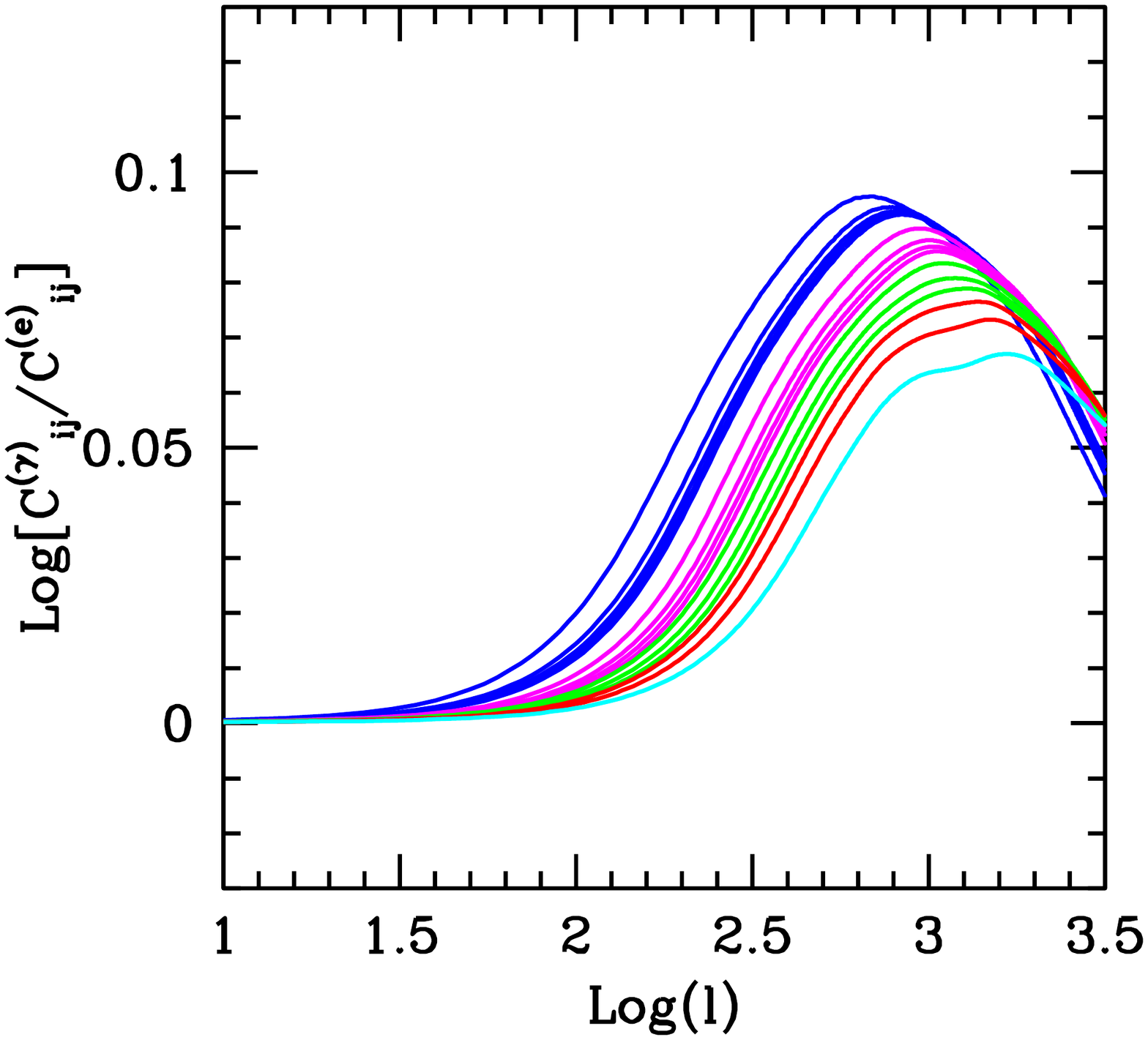} 
\caption{Ratio between spectra obtained from the 128$\gamma$ and
  128$e$ seeds, in the case of a 5 band tomography .}
 \label{R128-5}
\end{figure}

\begin{figure}
\epsscale{.9}
\plotone{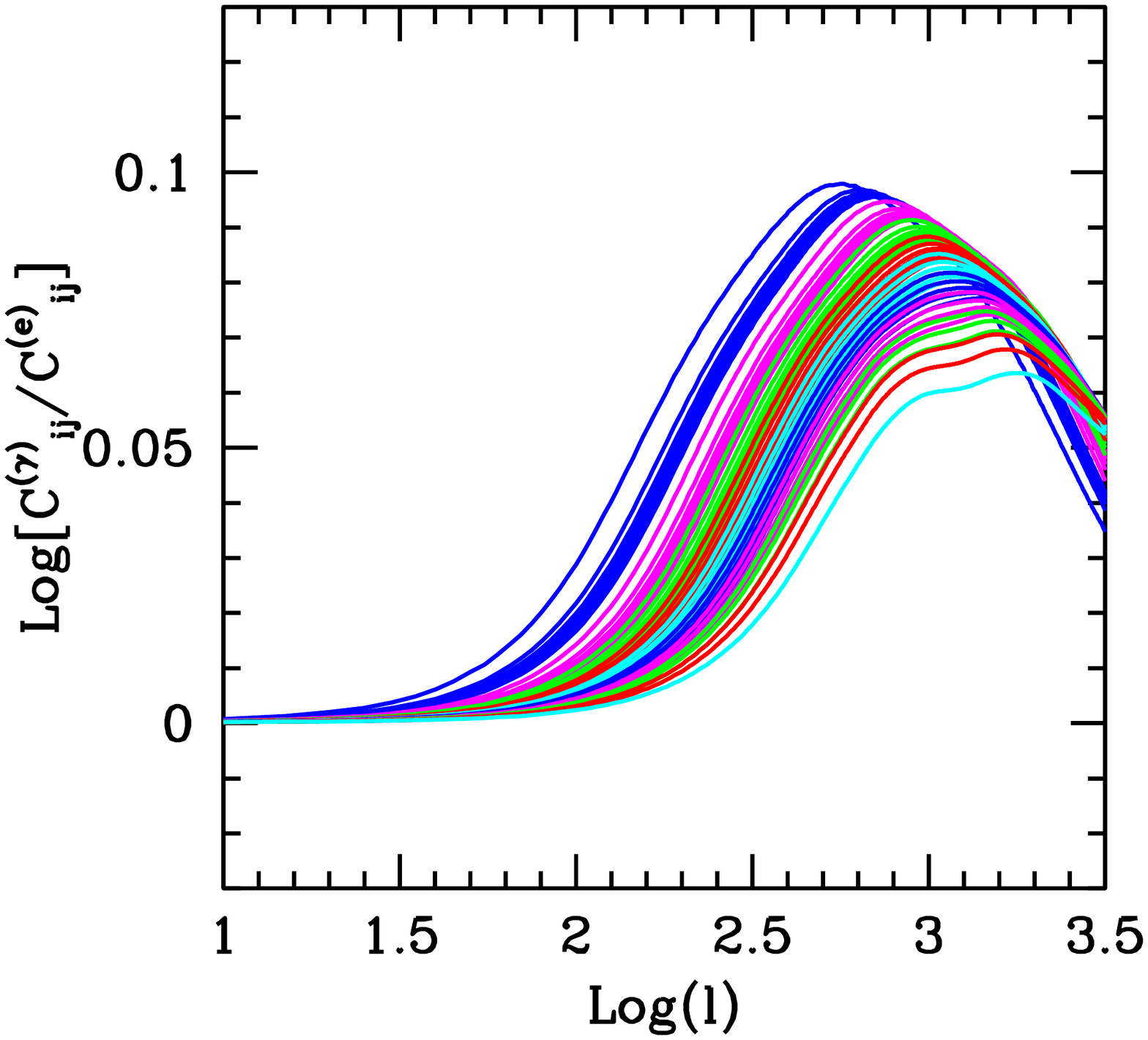} 
\vskip -3.truecm
\caption{As previous Figure for a 10 band tomography .}
 \label{R128-10}
\end{figure}

The discrepancy between seeds is better visible in Figures
\ref{R128-5} and \ref{R128-10}, for the 5 and 10 band cases,
respectively.
These plots allow us 2 comments: (i) The shear spectra discrepancy is
substantially independent from the number of bands and reaches $
24.7\, \%$ for $C_{11}$ for a 5 band tomography; this top discrepancy
is estimated similarly to $P_\delta$ spectra, by averaging over $\ell$
with $N=10$. (ii) If we remind that shear spectra are obtained by
integrating over redshift, so including contributions from $P_\delta$
up to $z \sim 2$, we appreciate that no substantial decrement of
spectral discrepancy occurs, when passing from fluctuation to shear
spectra.


\subsection{Shear spectra from simulations in the 512$\, h^{-1}$Mpc box }
\begin{figure}
\epsscale{.9}
\plotone{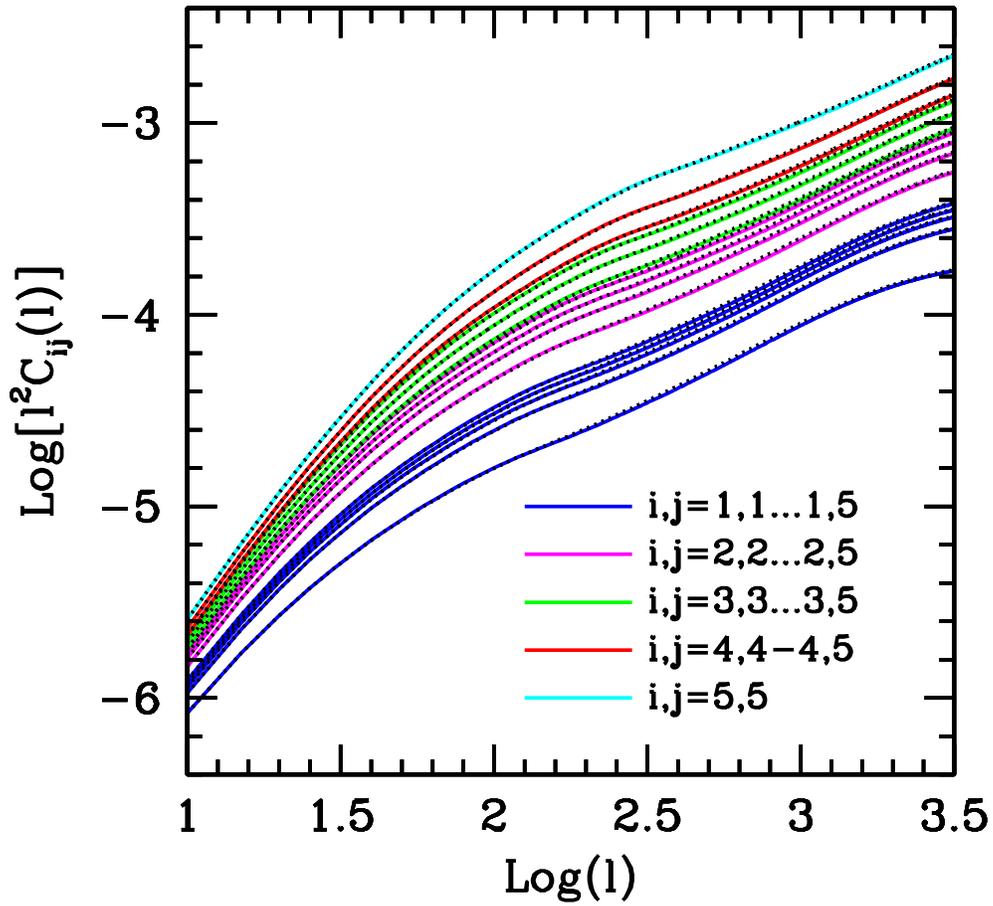}
\vskip -3.truecm
\caption{Shear spectra from the 512$\zeta(3)$ seed, in the case of a 5
  band tomography. Black dotted curves yield the spectra as obtainable
  from the 512$\sqrt{2}$ seed, however hardly distinguishable, in this
  case. }
 \label{512-5} 
\end{figure}

Discrepancies, as expected, are significantly smaller when a greater
simulation box is used. In Figure \ref{512-5} we plot shear spectra
obtained from the 512$\zeta(3)$ seed, in the cases of a 5 band
tomography. As in the smaller box case, we overlap these spectra with
those obtainable from the 512$\sqrt{2}$ seed, which is the most
distant from average and, in this case, exceeds average.
Discrepancies, however, are quite hard to be perceived in this way.

\begin{figure}
\epsscale{.9}
\plotone{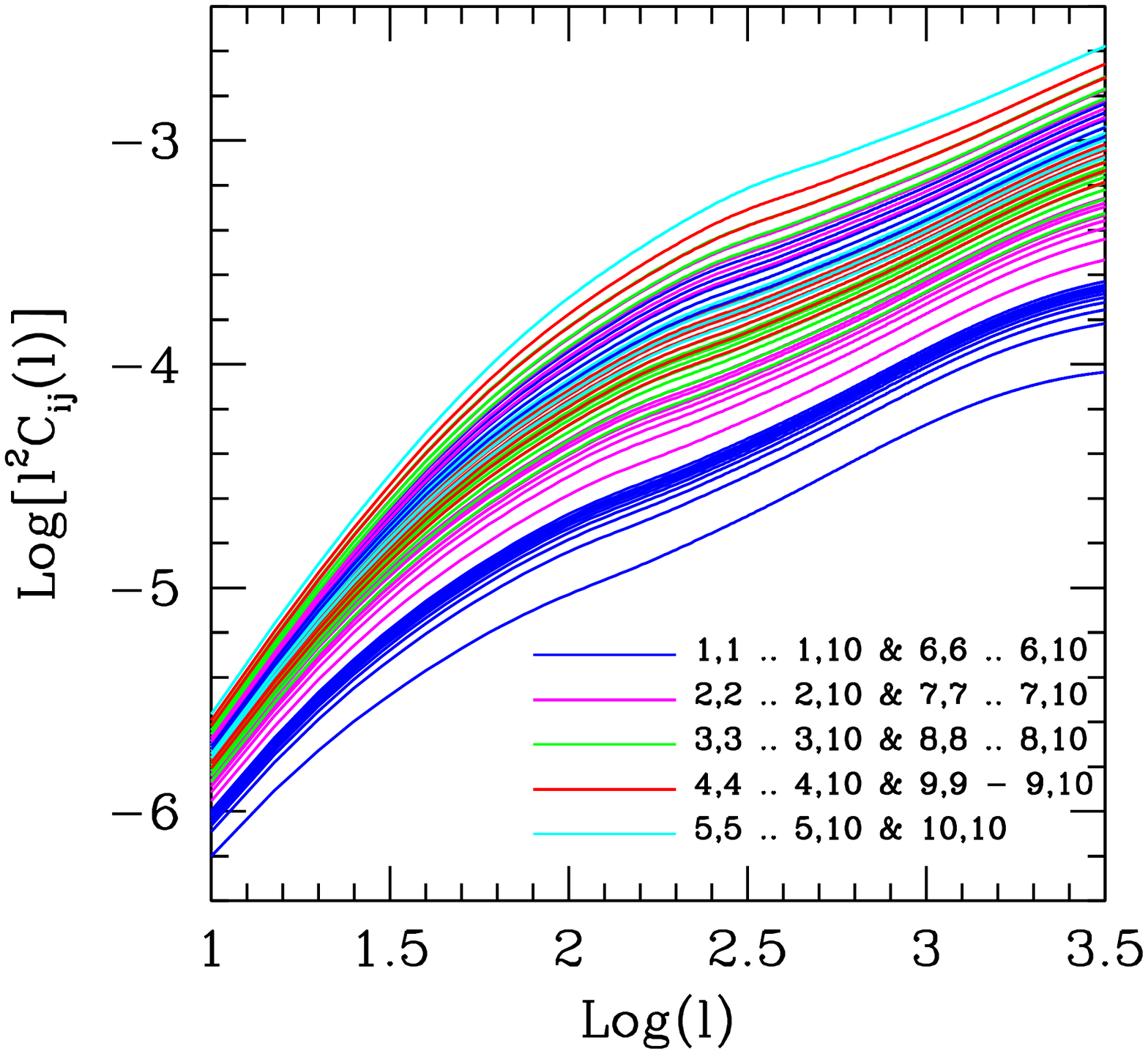}
\vskip -3.truecm
\caption{Shear spectra from the 512$\sqrt{2}$ seed, in the case of a
  10 band tomography. }
 \label{512-10} 
\end{figure}

In Figure \ref{512-10} we then plot shear spectra in the case of a 10
band tomography, as obtainable from the 512$\sqrt{2}$ seed.

\begin{figure}
\epsscale{.9}
\plotone{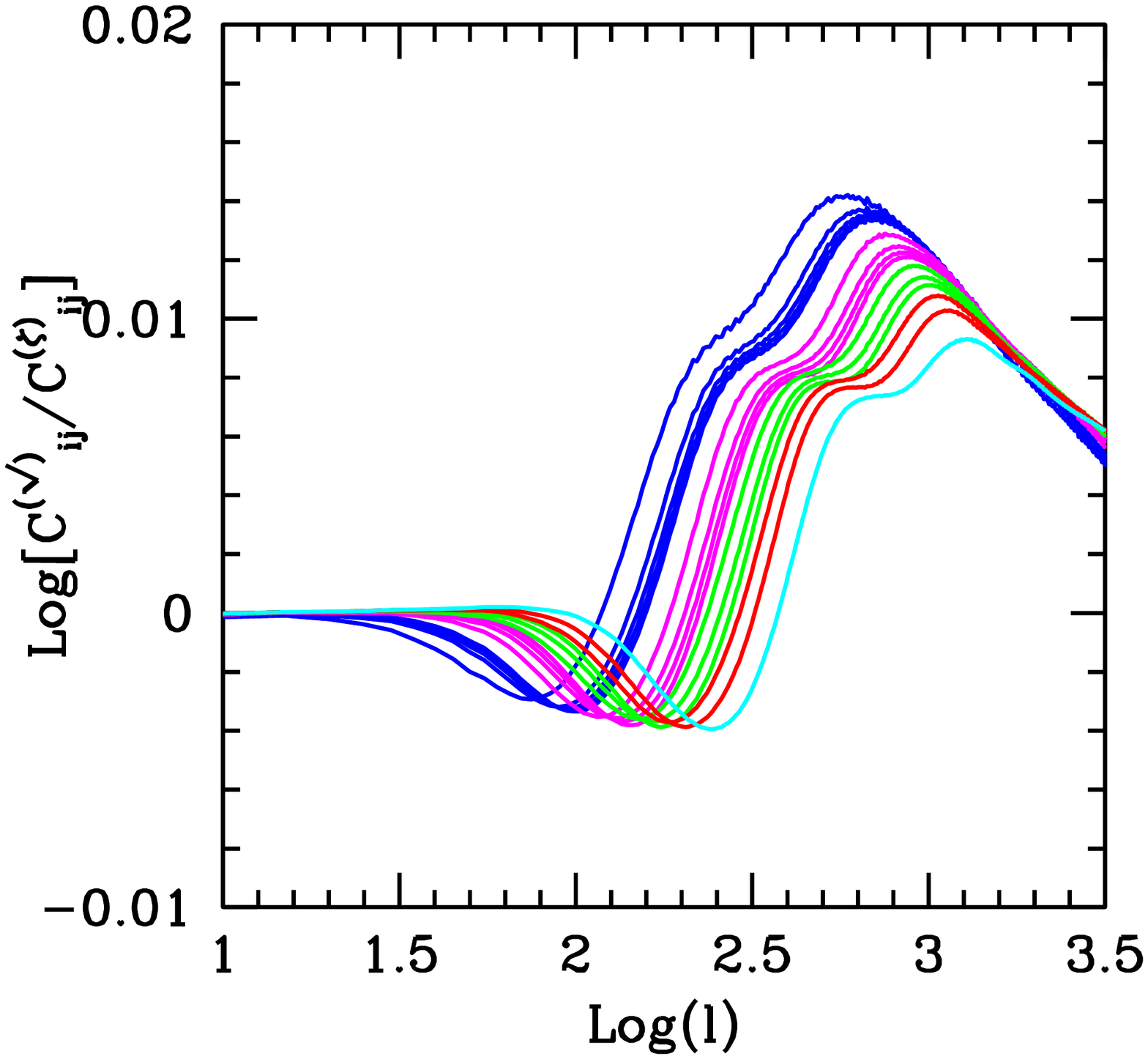} 
\vskip -3.truecm
\caption{Ratio between spectra obtained from the 512$\sqrt{3}$ and
  512$\zeta(3)$ seeds, in the case of a 5 band tomography .}
 \label{R512-5}
\end{figure}

\begin{figure}
\epsscale{.9}
\plotone{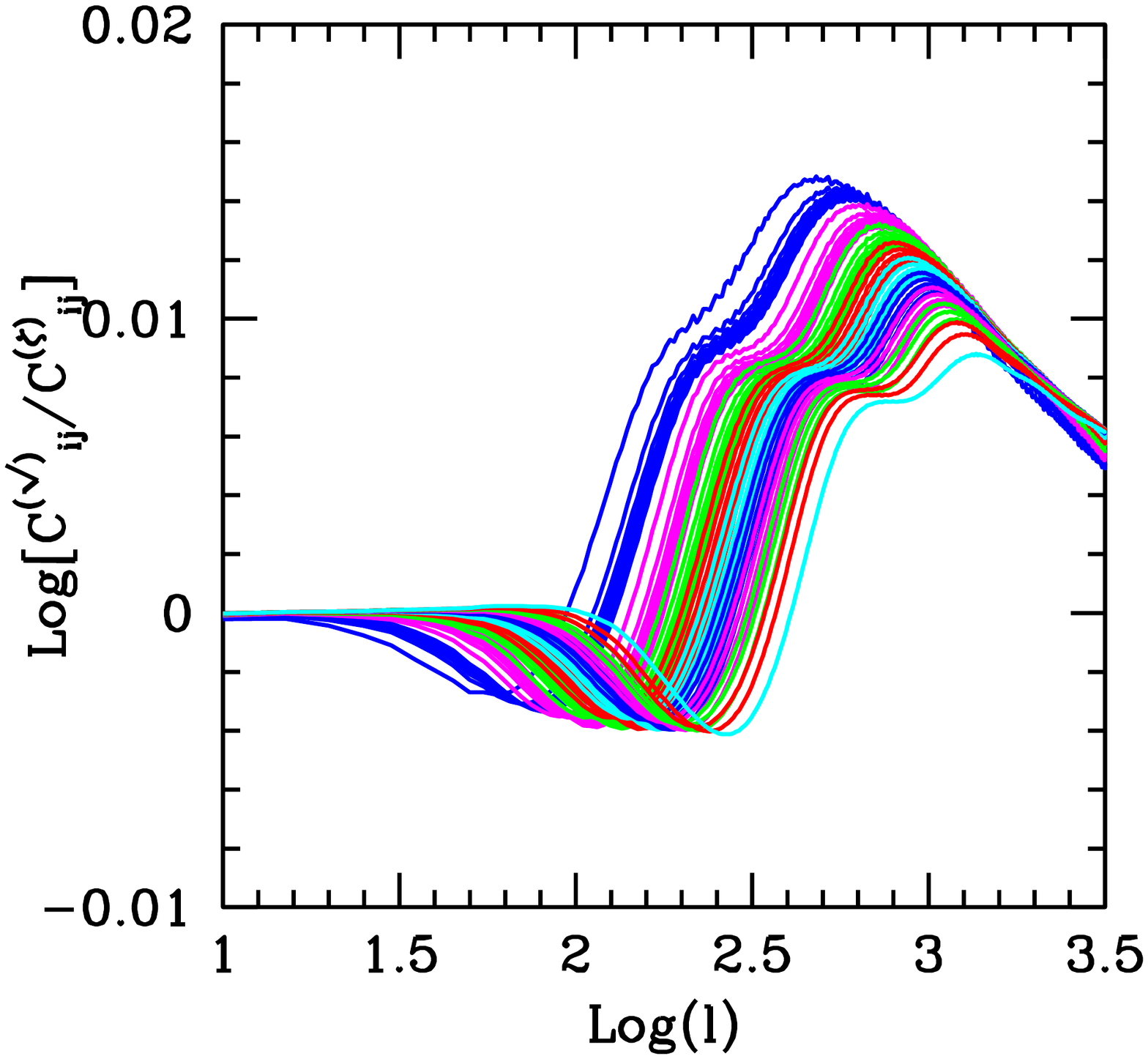} 
\vskip -3.truecm
\caption{As previous Figure, in the case of a 10 band tomography .}
 \label{R512-10}
\end{figure}

In Figures \ref{R512-5} and \ref{R512-10}, we finally plot the ratios
between the shear spectra obtained from the simulations 512$\zeta(3)$
and 512$\sqrt{2}$ in the cases of 5 and 10 bands.


Tiny oscillations in the curve arise from residual oscillations of
$P_\delta$, visible here because of the reduced amplitude of the
overall ratio; when comparing Figures \ref{R512-5} \& \ref{R512-10}
with Figures \ref{R128-5} \& \ref{R128-10}, take also notice of the
reduced range of the ordinates.

Once again we appreciate that discrepancies are almost independent
from the number of tomographic bands and their top value is $\sim
3.32\, \%$.

Let us finally outline the peculiar feature that ratios exhibit at low
$\ell$, where they become less than unity. As a matter of fact, at low
$k$, the very ratio between $P_\delta$ spectra exhibits significant
oscillations about unity (see Figure \ref{Pratio1}), smeared out by
the integrals (\ref{pijl}), yielding the prevealing contribution, in a
direction opposite to the main trend at greater $k$'s.

This kind of peculiarities become more and more significant as we go
to greater boxes and, consequently, also to smaller bulk
discrepancies. If making use of different seeds, for which such bulk
discrepancy is however smaller, we find similar anomalies. The
essential point is that they fall in a region necessarily separate
from that allowing us the basic estimate of top discrepancies.

\section{Discussion and conclusions}
This paper makes use of sets of $\Lambda$CDM N--body simulations to
investigate sample variance, in view of the possible use of {\it
  ad--hoc} simulations to fit data exhibiting evident deviation from
GR or other peculiar effects due to the nature of the dark components.

Such evidence could derive from the analysis of weak lensing data.
Accordingly, the effects of sample variance on tomographic shear
spectra are also inspected.

Our analysis is based on 8 $N$--body simulations in boxes of
increasing side $L$. According to simple probabilistic arguments, the
spread between their results approaches $\sim 1.5\, \sigma$'s.
Experiments in progress plan to achieve a precision $\sim 1\, \%$;
accordingly, the size of the simulation box and the related resolution
should enable us to reach the same precision, at least.
\begin{figure}
\epsscale{.8}
\plotone{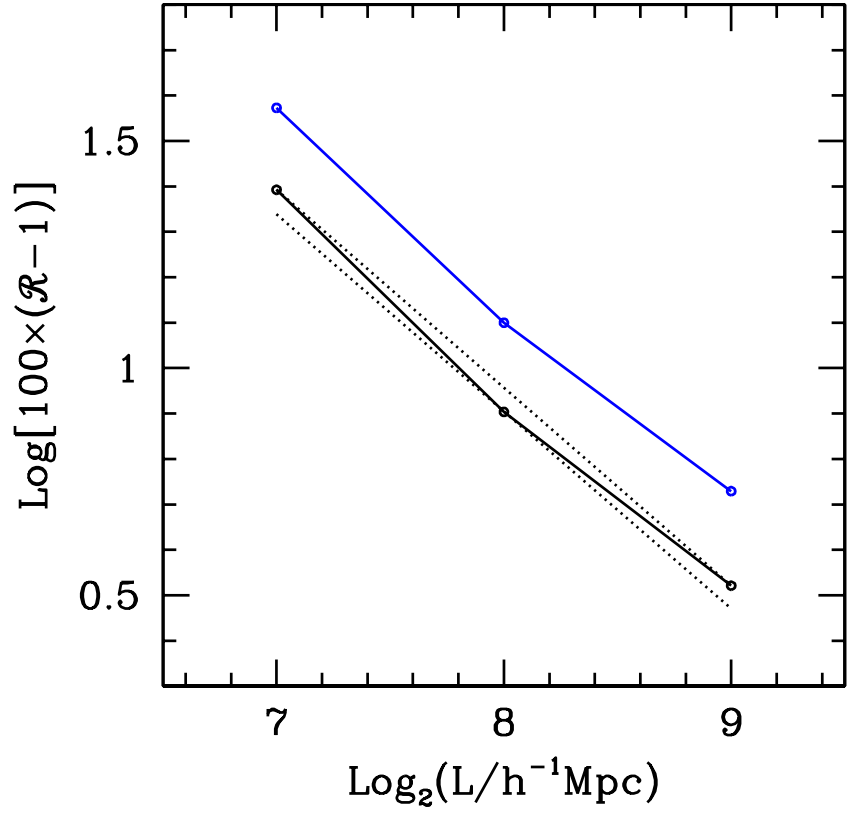}
\caption{Top ratio ${\cal R}_{max}$, between fluctuation or shear
  spectra (in blue or black, respectively). In the latter case values
  are for a 10 band tomography, although the dependence on the band
  number is almost negligible. \label{resol} }
\end{figure}

In Figure \ref{resol} we then plot the maximum discrepancies 
  between the ``most distant'' realizations considered as a function
of the box side.  The $k$ values where they arise lay in the
  spectral region where $k$ discreteness is still relevant. It is so,
  in spite of their being damped by our procedure, using sets of
  simulations points to ``bend'' linear spectra, so to account for the
  rise of non--linearity.  More precisely, in Figure \ref{resol}, we
consider the ratios ${\cal R} = \left|
\log[P_\delta^{(1)}(z=0)]/P_\delta^{(2)}(z=0)] \right| $ (in blue) or
  $\left| \log(C_{ij}^{(1)}/C_{ij}^{(2)}) \right| $ (in black); apices
  refer to the 2 seeds considered (closest and farthest from average)
  for each box side.
Such ratio overcome the unity by an amount which, multiplied by 100,
yields the percent discrepancy.  This surely yields values
  greater than an estimate referring to spectral covariance, based on
  averaging among discrepancies, as was done by previous authors (see,
  in particular, \cite{Takahashi}). In Figure \ref{resol} we show
that the percent discrepancy between the most distant realizations
decreases almost linearly with the box side, being cut by a factor
$\sim 0.35 $--0.45 when the side doubles. This is true both for
$P_\delta(k)$ and $C_{ij}(\ell)$.


 The decrease is surely expected, as the number of realizations
  considered within each box increases with the box size. An analogous
  decrease is shown, e.g., in Fig.~(12) of \cite{Takahashi}, although
  they plot spectral covariance and their increase in the realization
  number arises from increasing the number of equal size boxes taken,
  within a very large number of independent simulation boxes.

Being wider, however, the size of our dependence on the box side is
safer, in respect to varying I.C. in a set of 8 simulation boxes. An
estimate of such residual dependence is obtainable from the apparent
deviation of the $L$ dependence from linear. In Figure \ref{resol} and
for the $C_{ij}$ discrepancies, two dashed lines frame the expected
interval for a linear behavior. If we assume that the uncertainty of
each estimate is set by the width of such interval, we find $\sim 13\,
\%$.  Similarly, for $P(k)$, we find $\sim 15\, \%$.

If we extrapolate the linear behavior to seek where realization
discrepancies within a 8 box set, defined according to our criterion,
are expected to lay below $1\, \%$, we find a box side in the range
$1300$--$1700\, h^{-1}$Mpc, within the allowed range of slopes.   It
is also clear that discrepancy estimates do not depend just on box
size, but also on mass and force resolution. Our choice is however a
typical one, meant to optimize the results of the numerical effort.

This lead us to conclude that a set of 8 simulations in a box with
512$\, h^{-1}$Mpc aside is still insufficient to provide a unbiased
test for any model, if the precision level wanted ranges around $1\,
\%~;$ such aim might be however approached with a box side $\sim 3$--4
times greater,  keeping at the same resolution level and,
  therefore, accordingly increasing the dynamical range.
  

 As many reseachers previously did, e.g. to provide spectral
  estimators (see, e.g., \cite{Heitmann et al. 2010, Heitmann et
    al. 2013, Lawrence}), one could prefere to run a large number of
  simulations in smaller boxes, aiming to examine the same number of
  realizations without expanding so far the required dynamical
  range. For instance, one could replace the 8 boxes $ \sim 1700\,
  h^{-1}$Mpc aside, with $\sim 300$ boxes with $\sim 500\, h^{-1}$Mpc
  aside or $\sim 60$--70 realizations in boxes with $\sim 800\,
  h^{-1}$Mpc aside.

A first point to outline is then that including the impact of DC
modes, i.e., of wavelength wider than the simulation box, marginally
irrelevant when the statistics is limited to 8 boxes, would then
become indispensable.

Changing the box size, however, bears a further consequence, as the
density of values along the $k$ axis is bound to vary. Changing the
discrete mode set in the simulation will then affect the covariance
of the 3--D mass density. In turn, this affects the covariance of the
mass density power spectrum. These mode--coupling effects were first
outlined by \cite{Hamilton}
(see also recent results by \cite{Takada} and \cite{Li}).

Sample variance is directly connected to that, so that we cannot
restrict ourselves to barely counting the number of realizations as a
function of the overall volume. This point is to be borne in mind and
carefully weighted when a further effort to attain $C_{ij}$ prediction
at precision level $\cal O$$(1\, \%)$ were deployed.

\vskip .3truecm

\noindent
{\bf 
Acknowledgments}

An anonymous referee is to be thanked, in particular for outlining us
the effects on covariance of changing the box size. SAB acknowledges
the support of the Italian CIFS. LC and OFP are grateful to CNPq
(Brazil) and Fapes (Brazil) for partial financial support. This work
has made use of the computing facilities of the Laboratory of
Astroinformatics (IAG/USP, NAT/Unicsul), whose purchase was made
possible by the Brazilian agency FAPESP (2009/54006-4) and the INCT-A.

\end{document}